\documentclass[12pt]{article}
\usepackage{latexsym}
\usepackage{epsf}
\usepackage{hyperref}

\def\NON{\nonumber\\}
\def\NXT{\\}
\def\bibi{\bibitem}

\def\a{\alpha}
\def\b{\beta}

\def\d{\delta}
\def\e{\epsilon}                
\def\g{\gamma}

\def\k{\kappa}

\def\m{\mu}

\def\p{\pi}                     
\def\r{\rho}                    

\def\P{\Pi}

\def\ca{{\cal A}}
\def\cb{{\cal B}}
\def\cc{{\cal C}}
\def\cd{{\cal D}}

\def\cf{{\cal F}}
\def\cg{{\cal G}}
\def\ch{{\cal H}}   

\def\ck{{\cal K}}
\def\cl{{\cal L}}
\def\cm{{\cal M}}

\def\co{{\cal O}}

\def\cs{{\cal S}}

\def\cv{{\cal V}}
\def\cw{{\cal W}}

\def\bo{\raisebox{-.4ex}{\large$\Box$}}                 
\def\cbo{{\,\raise-.15ex\Sc [\,}}                       
\def\dg{^\dagger}                                     

\def\vev#1{\Big\langle #1 \Big\rangle}           

\def\svev#1{\left\langle #1\right\rangle}       

\def\ddt#1{{\buildrel {\hbox{\LARGE .\kern-2pt.}} \over {#1}}}

\def\secteq#1{ \setcounter{equation}{0}
               \renewcommand{\theequation}{#1.\arabic{equation}} }

\def\sstyle{\scriptstyle}

\def\ie{\mbox{\it i.e.} }
\def\eg{\mbox{\it e.g.} }

\def\frac#1#2{ {\sstyle {#1\over #2} } }
\def\det#1{{\rm det}\left(#1\right)}

\def\tr{{\rm tr}\,}

\def\half{{1\over 2}}

\def\det{{\rm det\,}}

\thicklines                         
\def\Cbar{{\overline{C}}}
\def\cbar{{\overline{c}}}
\def\sbar{{\overline{s}}}
\def\bbar{{\overline{b}}}
\def\qbar{{\overline{q}}}
\def\ubar{{\overline{u}}}
\def\dbar{{\overline{d}}}
\def\Xbar{{\overline{X}}}
\def\chibar{{\overline{\chi}}}
\def\psibar{{\overline{\psi}}}
\def\Xt{{\tilde{X}}}
\def\St{{\tilde{S}}}
\def\Pt{{\tilde{P}}}
\def\shat{{\hat{s}}}
\def\khat{{\hat{k}}}
\def\kvec{{\vec{k}}}

\def\dg{{\delta g}}

\def\eg{{\it e.g.\ }}
\def\ie{{\it i.e.}}
\def\cfdot{{\it cf.}}

\def\OUT#1{}

\begin{document}
\hyphenation{fer-mio-nic per-tur-ba-tive pa-ra-me-tri-za-tion
pa-ra-me-tri-zed a-nom-al-ous}

\begin{center}
\vspace{10mm}
{\large\bf $SU(N)$ chiral gauge theories on the lattice}
\\[12mm]
Maarten Golterman$^a$\ \ and \ \ Yigal Shamir$^b$
\\[8mm]
{\small\it
$^a$Department of Physics and Astronomy,
San Francisco State University\\
San Francisco, CA 94132, USA}\\
{\tt maarten@stars.sfsu.edu}
\\[5mm]
{\small\it $^b$School of Physics and Astronomy\\
Raymond and Beverly Sackler Faculty of Exact Sciences\\
Tel-Aviv University, Ramat~Aviv,~69978~ISRAEL}\\
{\tt shamir@post.tau.ac.il}
\\[10mm]
{ABSTRACT}
\\[2mm]
\end{center}

\begin{quotation}
We extend the construction of lattice chiral gauge theories based
on non-perturbative gauge fixing to the
non-abelian case.  A key ingredient is that fermion doublers
can be avoided at a novel type of critical point which is
only accessible through
gauge fixing, as we have shown before in the abelian case.
The new ingredient allowing us to deal with the non-abelian
case as well is the use of equivariant gauge fixing,
which handles Gribov copies correctly, and avoids
Neuberger's no-go theorem.  We use this method in order
to gauge fix the non-abelian group (which we will take
to be $SU(N)$) down to its maximal abelian subgroup.
Obtaining an undoubled, chiral fermion content requires us to gauge-fix
also the remaining abelian gauge symmetry.
This modifies the equivariant BRST identities,
but their use in proving unitarity remains
intact, as we show in perturbation theory.
On the lattice, equivariant BRST symmetry as well
as the abelian gauge invariance are broken, and a judiciously chosen
irrelevant term must be added to the lattice gauge-fixing
action in order to have access to the desired critical point
in the phase diagram.
We argue that gauge invariance is restored in the continuum limit
by adjusting a finite number of counter terms.  We emphasize that
weak-coupling perturbation theory applies at the critical point
which defines the continuum limit of our lattice chiral gauge
theory.

\end{quotation}

\newpage
\vspace{5ex}
\noindent {\large\bf 1.~Introduction}
\secteq{1}
\vspace{3ex}

Attempts to develop a non-perturbative, lattice definition of
chiral gauge theories have a long history. To date,
no lattice definition of a non-abelian chiral gauge
theory which maintains exact gauge invariance is known.
(For abelian chiral gauge theories, see ref.~\cite{mlabelian}.)
The fundamental difficulty is that, even if the whole collection of
fermion fields is anomaly free, each lattice fermion field
needs to contribute its ``share" of the anomaly \cite{ks},
and the regulated theory therefore
tends to break the gauge invariance by irrelevant terms.

Because of those---classically, but not
quantum-mechanically---irrelevant couplings, the longitudinal
gauge degrees of freedom are {\it not} decoupled from the fermions.
Extensive studies that go back to the eighties and early nineties
have taught an important lesson:
The un-controlled non-perturbative dynamics of these unphysical degrees
of freedom tends to spoil the desired continuum limit
through the re-generation of doublers.
For reviews, see refs.~\cite{reviews,mg2000}.

A remedy is to regain control over the longitudinal dynamics by
non-perturbative gauge fixing. This idea was first proposed in ref.~\cite{rome}.
While the insight of ref.~\cite{rome} is an important one,
the proposal itself was incomplete.
In order that, indeed, a critical point will exist which describes the
gauge-fixed target continuum theory non-perturbatively,
non-trivial additional elements are needed,
as first introduced in refs.~\cite{sprd,gsplb}. In subsequent work, convincing
evidence was provided that fermion doublers are avoided, and that
this program can be carried out successfully
in abelian lattice chiral gauge theories [8--14].\footnote{
  This statement ignores triviality, the latter being a property of
  abelian gauge theories
  which is unrelated to the chirality of the fermion spectrum.
}

In this paper, we describe in detail a proposal for the
construction of $SU(N)$ chiral gauge theories on the lattice.
To make it clear at the outset what our proposal does and does
not accomplish, we begin with a summary of the main features
and open questions of our construction.

Given an asymptotically free
$SU(N)$  chiral gauge theory,\footnote{
  We believe that the extension to other groups is a purely technical matter.
} our method gives a prescription for how to discretize this theory in a
way that satisfies the following properties:

\begin{itemize}

\item The lattice theory is local.

\item  A straightforward, systematic weak-coupling expansion is valid near
the critical point at which the target continuum theory is defined
\cite{sprd,gsplb}.
Near this critical point the lattice theory is manifestly renormalizable
by power counting.

\item The fermions of the lattice theory are undoubled.
In other words, the chiral fermions of the formal target continuum
theory remain chiral on the lattice \cite{bgsprl,bgs97,bd}.

\item The fermion content has to be anomaly free in the usual sense.
The theory accounts correctly for fermion-number
violating processes \cite{fnv}.

\item In order to construct the lattice theory, the target
continuum theory is gauge-fixed first, before it is transcribed
to the lattice, in such a way as to
have access to a complete set of Slavnov--Taylor identities.  This is
a central element of our construction,
since our theory is not gauge invariant on the
lattice, and both Slavnov--Taylor identities as well as
power counting are needed in order to obtain a gauge-invariant
continuum limit with the adjustment of a finite number of
counter terms.  In this sense, our proposal follows the
philosophy of ref.~\cite{rome}.

\item In order to gauge fix the non-abelian ``part" of the
gauge symmetry, we employ a gauge-fixing method which may be regarded
as a variant of the maximal-abelian gauge, and is
known as equivariant gauge fixing \cite{schaden}.  This allows
us, as explained in detail in section 4, to circumvent the Gribov
problem in a rigorous manner.  In
particular, it allows us to put the well-known ``Faddeev--Popov
trick" on a rigorous footing, without running into
the impasse of Neuberger's theorem \cite{hnnogo}.

\end{itemize}

In a non-abelian theory without chiral fermions, equivariant gauge fixing
can be implemented non-perturbatively, while maintaining an
exact BRST-type invariance.  With chiral fermions, gauge
invariance is lost, since in order to avoid species doubling
we add a Wilson term \cite{smitswift}, which is not invariant under chiral
symmetry.  The gauge fixing is a key ingredient of our method:
it maintains power-counting renormalizability in the absence
of gauge invariance; it makes it possible to avoid doublers being generated
dynamically; and it allows us
to systematize the counter terms which need to be added to
obtain a gauge-invariant continuum limit where the unphysical
degrees of freedom decouple.

Of course, it is well known that for an arbitrary chiral
gauge theory obstructions exist to this program.
In the context of global (classical) chiral symmetry,
the triangle anomaly appears to be a fundamental reason for
the species-doubling phenomenon \cite{ks}.
In the context of local chiral invariance,
if the fermion spectrum is not gauge-anomaly free,
it will be impossible to recover gauge invariance (and, hence, unitarity)
in the continuum limit by tuning counter terms. Conversely, if the
fermion spectrum does satisfy the usual anomaly-cancellation condition,
one can recover gauge invariance and unitarity to all orders in
perturbation theory in the continuum limit \cite{suzuki,banetal}.

Other obstructions may
exist which prevent us from constructing the desired continuum limit.
A known example is the Witten anomaly \cite{witten,bc}.
Our construction does not answer the question of whether additional,
as yet unknown, non-perturbative obstructions exist, and if so, how
these depend on the gauge group and the fermion content.\footnote{
  We return to this issue in the conclusion section.
}  If a certain non-abelian chiral gauge theory {\it does} exist,
our construction provides a lattice
formulation of this theory in which
there is strong evidence supporting the existence of
a novel type of critical point yielding the correct
continuum limit. All the known pitfalls are avoided in our
construction.

At finite, non-zero lattice spacing our lattice definition of
a chiral gauge theory is not unitary, because of the presence
of unphysical states in the extended Hilbert space of a gauge-fixed
gauge theory and the lack of exact BRST invariance on the lattice.\footnote{
 For earlier work on unitarity in lattice chiral gauge theories, see
   ref.~\cite{smit}.
}
The existence of a systematic weak-coupling expansion in the
gauge coupling makes it possible to establish unitarity to
all orders in perturbation theory in the continuum limit.
Non-perturbative unitarity is an issue which will need to be investigated using
non-perturbative techniques.

To summarize, our construction does not prove the (non-perturbative)
existence of any unitary chiral gauge theories.  However, if a certain
chiral gauge theory does exist, our discretization of it should
be a valid discretization. Therefore our construction makes it possible
to systematically investigate fundamental questions pertaining to
asymptotically-free chiral gauge theories, using
non-perturbative techniques.

The underlying strategy is that,
keeping in mind that gauge invariance is broken explicitly on the lattice,
{\it one constructs a lattice theory containing a suitable gauge-fixing action
that admits a systematic perturbative expansion in the
coupling constant near the critical point where the continuum limit is taken.}
The existence of a systematic perturbative expansion
has the following implication: The elementary degrees of freedom
of the formal target (chiral) gauge theory, which may, by construction,
be identified from the classical continuum limit of the lattice action,
are indeed the elementary degrees of freedom obtained in the continuum
limit of the quantum theory. In particular, no doublers are generated,
and a chiral fermion spectrum can be maintained non-perturbatively.
The renormalized interactions also agree with those of the target continuum
theory, and all unphysical excitations decouple (at least)
to all orders in perturbation theory, after adding the appropriate
counter terms (as determined by the Slavnov--Taylor identities).
Standard power counting can be used in order
to organize the counter terms, which implies that there is a finite
number of them.

The construction of a theory with a critical point as alluded to above is
non-trivial.  The desired critical point exists thanks to the fact that the
gauge-fixing action on the lattice can be chosen such that
1) its unique absolute minimum is the configuration with all link variables
equal to the identity matrix \cite{gsplb}; 2) the lattice theory is manifestly
renormalizable by power counting (despite the fact that the regulated theory
is not gauge invariant), because it contains kinetic terms for all four
polarizations \cite{rome}.  As a result, the functional integral is
dominated by a single saddle point when the bare coupling $g_0$
is sufficiently small. This saddle point is controlled by a
straightforward weak-coupling expansion.  As usual,
the validity of lattice perturbation theory implies the existence of a
scaling region. In the case of an asymptotically free theory,
moreover, a physical infra-red scale will be dynamically generated
in the continuum limit $g_0 \to 0$. The critical point itself involves
this limit together with appropriate adjustment of the counter terms.

There are many ingredients to the construction of a theory with the
desired critical point, and most of these
have already been formulated and investigated in the past in the context
of abelian chiral gauge theory.
The absence of doublers has been established in rather much detail
\cite{bgsprl,bgs97,bd,bgspt}.  The global structure of the phase diagram
was studied in refs.~\cite{bgsphased,bglspd},\footnote{
  See also ref.~\cite{bds}.
}
which also contain
further tests of the validity of perturbation theory near the critical point,
and of the agreement between the light degrees of freedom
of the lattice theory and of the target continuum theory.
In this paper we will therefore only give a brief account of this program
(see Sect.~6), referring to earlier work for details.  In fact, for abelian
chiral gauge theories, no ghosts are needed if a linear gauge
is used, and, using an appropriate lattice transcription of the
Lorenz gauge, the construction of abelian lattice chiral gauge
theories was essentially completed before.

The remaining hurdle for non-abelian theories has little to
do with the fermions, and can first be addressed in the setting of
pure Yang--Mills theories.  The issue is the existence of
Gribov copies \cite{gribov}, and the problems which arise if
one tries to construct a path integral which sums over copies
with a correct weight.  There have been varying suggestions on
how to tackle this problem.  One idea is to sum over copies
with a measure that includes the Faddeev--Popov determinant
\cite{hirschfeld}.  Unfortunately, it was shown in a rigorous
setting that this does not work: Neuberger's theorem \cite{hnnogo}
asserts that the partition function of such a theory vanishes identically.
Another proposal with a
manifestly positive measure averages over gauge orbits through
a ``quenched'' scalar field \cite{pjlz}, but in this case we lack
a symmetry principle such as BRST in order to recover the
target continuum theory from the lattice \cite{bgos}.

Here, we develop the idea of ref.~\cite{schaden}, which can be used
to fix the gauge symmetry down to the maximal abelian subgroup,
evading Neuberger's theorem while maintaining an ``equivariant'' BRST
symmetry.  The remaining abelian gauge symmetry can then be fixed
without the need to introduce ghosts, much in the same way as in
our earlier work on abelian gauge fixing.  (This again avoids
Neuberger's theorem.)  The main content of
this paper then consists of two parts: first we explain how
to construct the equivariantly gauge-fixed lattice Yang--Mills theory;
and second, we explain how this construction may be adapted to accommodate
chiral fermions on the lattice.
The main goal in the first part (Sects.~2--5) is to show that
the equivariantly gauge-fixed lattice theory
is the same as the non-gauge-fixed theory.
The central issue in the second part (Sect.~6) is how to
gauge fix the remaining abelian gauge symmetry on the lattice,
and how to obtain the target
continuum theory after adding chiral fermions,
when the BRST symmetries of the target theory are broken in the
regulated theory.

This paper is organized as follows.  In Sect.~2, we discuss the
construction of an equivariantly gauge-fixed Yang--Mills theory
in the continuum, extending the results of ref.~\cite{schaden},
and establishing an extended BRST -- anti-BRST (equivariant) algebra
following ref.~\cite{baulieu};
we do the same on the lattice in Sect.~3.
We concentrate on the case of interest for $G=SU(N)$ lattice chiral gauge
theories, namely when the equivariant gauge fixing
leaves behind the local invariance under
the maximal abelian subgroup $H=U(1)^{N-1}$.
We construct a complete path integral with a local Boltzmann weight
which defines the equivariantly gauge-fixed lattice theory.
In addition to the gauge field, it contains ghost fields
taking values in the coset space $G/H$. In Sect.~4, we review
Neuberger's theorem, and prove rigorously that the equivariantly
gauge-fixed lattice theory is the same as the lattice theory
without any gauge fixing, thus evading the theorem.
By ``the same" we mean that, at finite lattice spacing,
correlation functions of gauge-invariant
operators are the same in both theories.

Of course, in order to develop (continuum or lattice) perturbation theory
for an equivariantly gauge-fixed Yang--Mills theory
with gauge group $G$, a ``second-stage'' gauge fixing
of the remaining subgroup $H \subset G$ will be required.
Since a complete gauge fixing in a renormalizable gauge
is needed for our goal, our fully gauge-fixed lattice theory also contains
a Lorenz gauge-fixing term for the remaining abelian subgroup.
Here several new issues arise. In Sect.~5, as a preparatory step,
we address them in the context of continuum perturbation theory,
again restricting ourselves to the case $G=SU(N)$, $H=U(1)^{N-1}$.
We introduce a yet larger BRST-type algebra involving a new,
abelian $H$-ghost sector. Since we are now (linearly) gauge fixing
an abelian symmetry, the new ghosts are {\it free} fields.
The equivariant BRST identities of the ``first-stage''
gauge fixing are modified, but we show that the complete
algebraic setup remains sufficiently potent to guarantee unitarity
(at least to all orders in perturbation theory).
We develop the relevant generalized BRST identities, and, employing these,
we work out a detailed example of how unitarity is maintained
in perturbation theory.

In Sect.~6 we finally turn to the construction of non-abelian lattice chiral
gauge theories.
The complete set of Slavnov--Taylor identities needed to establish gauge
invariance and unitarity of renormalized perturbation theory can evidently
be re-derived while omitting the (free!) abelian ghost terms from the
continuum action.
The target chiral gauge theory that we latticize is the {\it fully}
gauge-fixed continuum theory without the free abelian-ghost terms.
The definition of the lattice theory includes a lattice version of the
pure gauge action (for instance the plaquette action), a
chiral-fermion action (for instance that of Sect.~6),
the equivariant gauge-fixing action of Sect.~3,
a Lorenz gauge-fixing term for the remaining abelian subgroup,
an irrelevant term needed for the uniqueness of the classical vacuum,
and a counter-term action.
We review the mechanism that guarantees the
existence of the appropriate critical point,
which remains essentially the same as in our previous work
on abelian lattice chiral gauge theories.
Since the lattice theory is not gauge invariant,
we construct a complete set of counter terms.

We use the
concluding section for additional remarks and comments.
In particular, we compare our results with those of refs.~\cite{mlnonab,mlpert}
(which describe an attempt at an exactly gauge-invariant lattice formulation
of non-abelian chiral gauge theories),
and discuss future prospects.
A number of technical points is relegated to the four appendices.

\vspace{5ex}
\noindent {\large\bf 2.~Equivariantly gauge-fixed Yang--Mills theories
-- continuum}
\secteq{2}
\vspace{1ex}

In this section we will describe the equivariant gauge fixing of a
Yang--Mills theory.\footnote{
  For the case $G=SU(2)$, $H=U(1)$, see ref.~\cite{schaden}.
}
  Equivariant gauge fixing
fixes only part of the gauge group $G$, leaving a subgroup $H\subset G$
unfixed. The main result of this section is a continuum action invariant
under a set of BRST-type transformations satisfying an equivariant, extended
BRST -- anti-BRST algebra. (This action is also invariant
under a few related symmetries.)
Unless otherwise stated, the results of this section are valid for
any simple, compact $G$ and any (in general, not simple) subgroup $H\subset G$.
We work in euclidean space.
The Yang--Mills lagrangian with gauge coupling $g$ is
\begin{eqnarray}
\label{YM}
\cl_{YM}&=&{1\over 2g^2}\ \tr(F_{\mu\nu}^2)\,,\\
iF_{\mu\nu}&=&[D_\mu(V),D_\nu(V)]\,,\nonumber\\
D_\mu(V)&=&\partial_\mu+iV_\mu\,,\ \ \ \ \
V_\mu=V_\mu^a T^a\,,\nonumber
\end{eqnarray}
with $T^a$ the hermitian generators of $G$ normalized such that
$\tr(T^a T^b)=\frac{1}{2}\delta_{ab}$, and structure constants
$f_{abc}$ defined by $[T^a,T^b]=if_{abc}T^c$.  The structure constants
are fully anti-symmetric in all three indices.

We will now divide the generators into a subalgebra $T^i$ generating
the subgroup $H$, and the rest, $T^\alpha$, spanning the coset
space $G/H$.  Correspondingly, we write the vector field $V$
as
\begin{equation}
\label{AW}
V_\mu=V_\mu^a T^a=A_\mu^i T^i+W_\mu^\alpha T^\alpha\,.
\end{equation}
We will use indices $i,j,k,\dots$ to indicate $H$ generators,
and $\alpha,\beta,\gamma,\dots$ for generators in $G/H$.
We note that
\begin{equation}
\label{closure}
f_{i\alpha j}=-f_{\alpha ij}=-f_{ij\alpha}=0\,,
\end{equation}
because the product of two elements of $H$ should again be in $H$.
We choose a gauge-fixing condition which is covariant under $H$,
\begin{equation}
\label{gf}
\cf(V)=\cd_\mu(A)W_\mu\equiv\partial_\mu W_\mu+i[A_\mu,W_\mu]\,,
\end{equation}
where $\cd_\mu(A)$ is a covariant derivative with respect to $H$.
Denoting the algebras of the groups $G$ and $H$ by $\cg$ and $\ch$,
we introduce $\cg/\ch$-valued ghost fields
\begin{equation}
\label{ghosts}
C=C^\alpha T^\alpha\,,\ \ \ \ \ \Cbar=\Cbar^\alpha T^\alpha\,,
\end{equation}
along with a coset-valued auxiliary field $b=b^\alpha T^\alpha$,
and demand invariance of the gauge-fixed theory under
equivariant BRST (eBRST) transformations
\begin{eqnarray}
\label{ebrst}
sA_\mu&=&i[W_\mu,C]_\ch\,,\\
sW_\mu&=&\cd_\mu(A)C+i[W_\mu,C]_{\cg/\ch}\,,\nonumber\\
sC&=&\left(-iC^2\right)_{\cg/\ch}=-iC^2+X\,,\nonumber\\
s\Cbar&=&-ib\,,\nonumber\\
sb&=&[X,\Cbar]\,,\nonumber
\end{eqnarray}
in which
\begin{equation}
\label{x}
X\equiv\left(iC^2\right)_\ch=2iT^j\ \tr(C^2T^j)\,.
\end{equation}
The transformation rule for $C$ is similar to the standard
BRST case, but for the extra term $X$ which projects $sC$
back onto the coset space.  This modification affects the
nilpotency of eBRST transformations.  In fact, using that
$sX=0$,
\begin{equation}
\label{ssqC}
s^2C=-i[X,C]=\delta_XC\,.
\end{equation}
This does not vanish, but equals a gauge transformation
(denoted by $\delta_\omega$) in $H$ with parameter $\omega=X\in\ch$.
Requiring $s^2\Cbar=\delta_X\Cbar$ determines the eBRST transformation rule
for $b$, after which one verifies that $s^2 b=\delta_X b$ as well.
The second eBRST variation of any physical field follows from the
fact that the standard BRST transformation is nilpotent,
so that only the $X$ part in $sC$ leads to a non-vanishing
result,\footnote{For abelian $H$, $s^2A_\mu=\partial_\mu X$.}
\begin{eqnarray}
\label{ssqaw}
s^2A_\mu&=&\cd_\mu(A) X=\delta_XA_\mu\,,\\
s^2W_\mu&=&-i[X,W_\mu]=\delta_XW_\mu\,.\nonumber
\end{eqnarray}
These are again precisely gauge transformations in $H$,
proving that $s^2=\delta_X$ is equivariantly nilpotent.\footnote{
  It is easy to see
that, for the product of any two fields $\Phi_1$ and $\Phi_2$,
$s^2(\Phi_1\Phi_2)=(s^2\Phi_1)\Phi_2+\Phi_1s^2\Phi_2$.}

Following the standard approach, we would choose as a
gauge-fixing lagrangian
\begin{equation}
\label{lgfpartly}
\cl'_{gf}=s\,\tr\left(2\Cbar\cf+i\xi g^2\Cbar b\right)\,,
\end{equation}
in which $\xi$ is a parameter.
This lagrangian is
invariant under eBRST transformations, as follows from the
fact that it is invariant under $H$ transformations and from
equivariant nilpotency of $s$.
However, as we will see
next, this is not the most general possible gauge-fixing
lagrangian.

It is useful to introduce the concept of anti-eBRST
transformations, following ref.~\cite{baulieu}.  Denoting
anti-eBRST by $\sbar$, we have for the gauge fields
\begin{eqnarray}
\label{aebrstgf}
\sbar A_\mu&=&i[W_\mu,\Cbar]_\ch\,,\\
\sbar W_\mu&=&\cd_\mu(A)\Cbar+i[W_\mu,\Cbar]_{\cg/\ch}\,.\nonumber
\end{eqnarray}
Using this and a partial integration, the first term in
Eq.~(\ref{lgfpartly}) can be written as
\begin{equation}
\label{morse}
s\ \tr\left(2\Cbar\cf\right)=-s\sbar\ \tr(W^2)\,.
\end{equation}
It is clear that the ``pre-potential" $\tr(W^2)$ can be generalized
to any $H$-invariant, rotationally invariant dimension-2
operator with ghost number zero.  This allows us to add
a term proportional to $\tr(\Cbar C)$ and we will choose
\begin{equation}
\label{lgf}
\cl_{gf}=-s\sbar\ \tr\left(W^2+\xi g^2\Cbar C\right)\,.
\end{equation}
Obviously, in order to complete the definition of $\cl_{gf}$,
we have to specify $\sbar C$ and $\sbar\Cbar$.  We will turn to
this next.

The pre-potential in Eq.~(\ref{lgf}) is invariant under a discrete
flip symmetry\footnote{
  This symmetry is related to ``ghost hermiticity'' of ref.~\cite{baulieu}.
}
on the ghost fields, $F C=\Cbar$, $F\Cbar=-C$.
(We define $F\Phi=\Phi$ for all physical fields.)
This symmetry will be useful later on, and we will define
$\sbar$ in the ghost sector so as to have $\cl_{gf}$ be invariant
under this symmetry as well.
We thus define $\sbar$ in the ghost sector by applying a
flip transformation to Eq.~(\ref{ebrst}), \ie\
by demanding that $\sbar F({\rm field})=F s({\rm field})$.
In addition to Eq.~(\ref{aebrstgf}) this gives
\begin{eqnarray}
\label{aebrstghost}
\sbar\Cbar&=&\left(-i\Cbar^2\right)_{\cg/\ch}
=-i\Cbar^2+\Xbar\,,\\
\sbar C&=&i\bbar\,,\nonumber\\
\sbar\bbar&=&-[\Xbar,C]\,,\nonumber
\end{eqnarray}
where we (temporarily)
introduced a new field $\bbar\equiv F b$, and in which
\begin{equation}
\label{xbar}
\Xbar=F X=\left(i\Cbar^2\right)_\ch=2iT^j\ \tr(\Cbar^2T^j)\,.
\end{equation}
The field $\bbar$ is not independent if we require that
the $s$, $\sbar$ algebra closes on $H$.  While this can
be worked out on the gauge fields, it is easier to do it
on a matter field $\Phi$ in the fundamental representation
of $G$, for which
\begin{equation}
\label{ebrstmatter}
s\Phi=-iC\Phi\,,\ \ \ \ \ \sbar\Phi=-i\Cbar\Phi\,.
\end{equation}
One finds that
\begin{equation}
\label{anticomm}
\{s,\sbar\}\Phi=(-b+\bbar+\{\Cbar,C\})\Phi\,.
\end{equation}
Now setting
\begin{equation}
\label{bbar}
\bbar=b-\{\Cbar,C\}_{\cg/\ch}\,,
\end{equation}
Eq.~(\ref{anticomm}) becomes an $H$ transformation $\delta_\Xt\Phi$ with parameter
\begin{equation}
\label{xtilde}
\Xt=i\{\Cbar,C\}_\ch=2iT^j\ \tr(\{\Cbar,C\}T^j)\,.
\end{equation}
We thus end up with the extended eBRST algebra
\begin{equation}
\label{sbarsq}
s^2=\delta_X\,,\ \ \ \ \
\sbar^2=\delta_\Xbar\,,\ \ \ \ \ \{s,\sbar\}=\delta_\Xt\,.
\end{equation}
Note that $F\{\Cbar,C\}=-\{\Cbar,C\}$, so that $F\bbar=b$.
Eq.~(\ref{bbar}) can also be used to work out $\sbar b$,
using Eq.~(\ref{aebrstghost}).  Finally, we have for all fields
that
\begin{equation}
\label{flipssbar}
F s=\sbar F\,,\ \ \ \ \ F\sbar=-sF\,.
\end{equation}

The most general eBRST-invariant gauge-fixing lagrangian would
be any linear combination of $\cl'_{gf}$ and $\cl_{gf}$.
However, if we insist on flip symmetry, the only possible
choice is $\cl_{gf}$, which, in addition to flip symmetry,
eBRST symmetry, and $H$ gauge invariance,
also has anti-eBRST symmetry.  We note that
if the coset structure constants $f_{\alpha\beta\gamma}$ are
all equal to zero, there is no difference between the two cases,
because in that case $\sbar\ \tr(\Cbar C)=-i\ \tr(\Cbar b)$.\footnote{
  In general the second term in Eq.~(\ref{lgfpartly}) cannot be
  written as $\sbar({\rm anything})$.
  An example with $f_{\alpha\beta\gamma}=0$
  is $G=SU(2)$, $H=U(1)$ \cite{schaden}.
}
Our gauge-fixing lagrangian is thus
\begin{equation}
\label{lgfdetail}
\cl_{gf}=-s\sbar\ \tr\left(W^2+\xi g^2\Cbar C\right)
=\sbar s\ \tr\left(W^2+\xi g^2\Cbar C\right)\,,
\end{equation}
where
\begin{eqnarray}
\label{lgfwcc}
\sbar s\ \tr\left(W^2\right)
&=&-2\ \tr\left(\Cbar\cd_\mu(A)\cd_\mu(A)C\right)
+2\ \tr\left([W_\mu,\Cbar]_\ch[W_\mu,C]_\ch\right)\nonumber\\
&&-2i\ \tr\left(\Cbar\cd_\mu(A)[W_\mu,C]_{\cg/\ch}\right)
-2i\ \tr\left(b\cd_\mu(A)W_\mu\right)\,,\nonumber\\
\sbar s\ \tr\left(\Cbar C\right)
&=&
\tr(b^2)-\tr\left(b\{\Cbar,C\}\right)
+\tr\left(\left(\Cbar^2\right)_{\cg/\ch}\left(C^2\right)_{\cg/\ch}\right)
\nonumber\\
&&+\tr\left(\{\Cbar,C\}_{\cg/\ch}\right)^2
-\tr\left(\Xt^2\right)\,.
\end{eqnarray}
This lagrangian is invariant under flip symmetry.
This can be verified
directly from eq.~(\ref{lgfwcc}), but can also be seen as
follows.
{}From Eqs.~(\ref{sbarsq}, \ref{flipssbar})
it follows that $F s\sbar=\sbar F\sbar
=-\sbar sF=(s\sbar-\delta_\Xt)F$, and thus $s\sbar$
commutes with $F$ on any $H$-invariant expression.
Since
the pre-potential in Eq.~(\ref{lgfdetail})
is $H$ invariant, it follows that $\cl_{gf}$ is invariant
under flip symmetry.

We may integrate out the auxiliary field $b$ to arrive at the
form
\begin{eqnarray}
\label{lgfonshell}
\cl_{gf}&=&{1\over\xi g^2}\ \tr\left(\cd_\mu(A)W_\mu\right)^2\\
&&-2\ \tr\left(\Cbar\cd_\mu(A)\cd_\mu(A)C\right)
+2\ \tr\left([W_\mu,\Cbar]_\ch[W_\mu,C]_\ch\right)\nonumber\\
&&+i\ \tr\left((\cd_\mu(A)\Cbar)[W_\mu,C]+
[W_\mu,\Cbar](\cd_\mu(A)C)\right)\nonumber\\
&&+\xi g^2\Biggl(\tr\left(\left(\Cbar^2\right)_{\cg/\ch}
\left(C^2\right)_{\cg/\ch}\right)
+{3\over 4}\ \tr\left(\{\Cbar,C\}_{\cg/\ch}\right)^2
-\tr\left(\Xt^2\right)\Biggr)\,.\nonumber
\end{eqnarray}
Note that the ghosts' differential operator $\cm_{\alpha\beta}$, defined
by the part bilinear in $\Cbar_\alpha$ and $C_\beta$, is self-adjoint and
real, and therefore symmetric.  The on-shell eBRST and anti-eBRST
transformation rules for $\Cbar$ and $C$ can be derived as usual from the
equation of motion for $b$.  The quartic ghost interactions are a
novel feature, and will play an important role in Sect.~4 below.

In this paper we are mainly interested in the application
to chiral lattice gauge theories, where we will take $G=SU(N)$ and
$H=U(1)^{N-1}$, the maximal abelian subgroup of $SU(N)$.
In this special case, even though the gauge-fixing terms break the original
$SU(N)$ gauge symmetry (even the global group), there is a
discrete subgroup which, in addition to the maximal abelian subgroup $H$,
remains a symmetry
of the full gauge-fixed action.  We first define this group in the
fundamental, $N$-dimensional representation.  Since now $H$ is
abelian, we can choose all generators $T^i$ of $H$ to be diagonal,\footnote{
  For instance, for $SU(2)$, $T^i\in\{\sigma_3/2\}$
  with $\sigma_k$ the Pauli matrices,
  and for $SU(3)$, $T^i\in\{\lambda_3/2,\lambda_8/2\}$,
  with $\lambda_a$ the Gell-Mann matrices.
}
while the remaining generators $T^\alpha$ are off-diagonal.
They may be written as
\begin{equation}
\label{offdg}
T^\alpha\to T^k_{AB}\,,\ \ \ \ \ k=1,2\,,\ \ \ \ \ 1\le A<B \le N\,,
\end{equation}
where $T^k_{AB}$, $k=1,2,3$, is
defined by the requirement that if we keep only the $A$-th and $B$-th
row and column, this matrix reduces to
${1\over 2}\sigma_k$
with all other entries of $T^k_{AB}$ being zero.
The ``skewed" permutation group $\St_N$ is defined as the subgroup
of $SU(N)$ generated by the elements
\begin{equation}
\label{snel}
\Pt^k_{(AB)}=\exp{(i\pi T^k_{AB})}\,,\ \ \ \ \ k=1,2\,.
\end{equation}
Acting on a vector of length $N$, this permutes the $A$-th and
$B$-th entries, and multiplies them by a factor $\pm 1$ or $\pm i$,
while leaving the other entries unchanged.
The discrete group $\St_N$ contains in particular the elements $\Pt^3_{(AB)}$
which also belong to $H$. On gauge fields and ghost-sector fields,
the action of $\Pt^k_{(AB)}$ is defined by
\begin{eqnarray}
\label{snadj}
V_\mu&\to&\Pt^k_{(AB)}V_\mu\left(\Pt^k_{(AB)}\right)^\dagger\,,\\
C&\to&\Pt^k_{(AB)}C\left(\Pt^k_{(AB)}\right)^\dagger\,,\nonumber
\end{eqnarray}
and likewise for $\Cbar$ and $b$.  It is straightforward to check
that $V_\mu=A_\mu+W_\mu$ does not transform in an irreducible
representation of $\St_N$,
but that $A_\mu$ and $W_\mu$ each transform separately
(and irreducibly).
On the generators of $H$ we have that
\begin{equation}
\label{hgen}
\left(\Pt^k_{(AB)}\right)^\dagger T^i\Pt^k_{(AB)}=R_{(AB)}^{ij}T^j\,,
\end{equation}
with $R_{(AB)}$ an $(N-1)\times(N-1)$ orthogonal matrix.  Note that
on the $T^i$, the group $\St_N$ just permutes the $A$-th and $B$-th
diagonal elements of each $T^i$.  $R_{(AB)}$ is thus independent
of $k$ being $1$ or $2$.
The pre-potential in Eq.~(\ref{lgfdetail}) clearly is invariant
under $\St_N$, and thus our equivariantly gauge-fixed Yang--Mills
theory is.

Finally, it turns out that the gauge-fixing action~(\ref{lgfdetail}) is invariant
under an $SU(2)$ group that acts on the ghost fields $C$ and $\Cbar$.
We will refer to this symmetry as ghost-$SU(2)$.\footnote{
  This symmetry exists for any $G$ and $H$.
  In ref.~\cite{schaden} it is referred to as $SL(2,R)$ symmetry.
}
The three generators are
\begin{eqnarray}
  \P_+ &=& C_\a\, {\d\over \d \Cbar_\a} \,, \qquad
  \P_- = \Cbar_\a\, {\d\over \d C_\a} \,,
\label{Ppm}
\NXT
  \P_3
  &=& C_\a {\d\over \d C_\a} - \Cbar_\a {\d\over \d \Cbar_\a} \,.
\label{P3}
\end{eqnarray}
$\P_3$ is the ghost-number charge.
These generators satisfy the same commutation relations
as $\sigma_\pm=\half(\sigma_1\pm i\sigma_2)$
and $\sigma_3$. Under ghost-$SU(2)$,
the ghost fields transform as a doublet $(C,\Cbar)$.

The action is evidently invariant under the ghost-number symmetry.
Let us establish its invariance under the extended,  ghost-$SU(2)$ symmetry.
The invariance of the terms bilinear in the ghost fields follows,
as in the case of flip symmetry, from the fact that the
operator $\cm_{\alpha\beta}$ is symmetric.
Turning to the on-shell four-ghost action,  we may rewrite it as
\begin{eqnarray}
  (\xi g^2)^{-1}\,\cl^{(4)}_{gh}
  &=&
  \tr\!\left( -\Cbar^2 C^2 -{1\over 4}\, \{\Cbar,C\}^2
    +\Xbar X -{1\over 4}\, \Xt^2\right)
\NON
  &=&
  -{1\over 3}\tr\!\left(\Cbar^2 C^2 -{1\over 4}\, \{\Cbar,C\}^2\right)
    +\tr\!\left(\Xbar X -{1\over 4}\, \Xt^2\right) \,.
\label{gh4b}
\end{eqnarray}
Each trace on the last row is invariant under ghost-$SU(2)$.
We used (anti)cyclicity of the trace
of the product of four ghost fields, from which it follows that
$\tr(\Cbar^2 C^2) = -\half \tr(\{\Cbar,C\}^2)$.
In the off-shell formalism,
the auxiliary field transforms under ghost-$SU(2)$ as
$\delta b_\alpha
= {i\over 2} \delta (f_{\alpha\beta\gamma} \Cbar_\beta C_\gamma)$.

\vspace{5ex}
\noindent {\large\bf 3.~Equivariantly gauge-fixed Yang--Mills theories
-- lattice}
\secteq{3}
\vspace{3ex}

In this section we show how the continuum theory constructed in the
previous section can be transcribed
to the lattice without any loss of symmetries (except rotational symmetry).
We limit the discussion to $G=SU(N)$, $H=U(1)^{N-1}$ from now on.\footnote{
  There exist straightforward generalizations of the lattice action(s)
  to other subgroups. For $N=2$, see ref.~\cite{schaden}.
}
First, the eBRST and anti-eBRST transformations
of the gauge field $V_\mu=A_\mu+W_\mu$ can be summarized as
\begin{equation}
\label{ebrstv}
sV_\mu=D_\mu(V)C\,,\ \ \ \ \ \sbar V_\mu=D_\mu(V)\Cbar\,,
\end{equation}
where the derivative is covariant with respect to the
full gauge group $G$, $D_\mu(V)C=\partial_\mu C
+i[V_\mu,C]$ (\cfdot\ eq.~(\ref{YM})).  The lattice version of these transformation
rules is, with $U_{x,\mu}=\exp(iV_\mu(x))$,
\begin{equation}
\label{ebrstlattice}
sU_{x,\mu}=i(U_{x,\mu}C_{x+\mu}-C_xU_{x,\mu})\,,
\ \ \ \ \ \sbar U_{x,\mu}=i(U_{x,\mu}\Cbar_{x+\mu}-\Cbar_xU_{x,\mu})\,.
\end{equation}
For the Yang--Mills lagrangian, Eq.~(\ref{YM}), we will assume
the usual plaquette action.  For the gauge-fixing action, we
choose the lattice version
\begin{eqnarray}
\label{lgflattice}
\cl_{gf}^L&=&-s\sbar\sum_\mu\sum_i\left(
-2\ \tr\left(T^iU_{x,\mu}T^iU^\dagger_{x,\mu}\right)
+\xi g^2\ \tr(\Cbar_xC_x)\right)\\
&=&-s\sbar\left({1\over 2}\sum_\mu\sum_\alpha
W_{x\mu}^\alpha W_{x,\mu}^\alpha
+\xi g^2\ \tr(\Cbar_xC_x)+O(V^4)\right)\,,\nonumber
\end{eqnarray}
which is $H$ invariant (recall that we take $H$ to be abelian).
In the second line, we used that
\begin{equation}
\label{ffd}
\sum_{i\gamma}
f_{i\gamma\alpha}f_{i\gamma\beta}=\delta_{\alpha\beta}\,.
\end{equation}

For the lattice construction of chiral gauge theories, or
in order to develop weak-coupling perturbation theory,
the remaining abelian gauge symmetry ($H$) will also need
to be fixed.  We will return to this in the sections to follow.

We will now work out the lattice equivalent of Eq.~(\ref{lgfonshell}).
First define
\begin{equation}
\label{cw}
\cw_{x,\mu}=-i\sum_i[U_{x,\mu}T^iU^\dagger_{x,\mu},T^i]
=W_{x,\mu}+O(V^2)\,,
\end{equation}
and lattice covariant derivatives $D^\pm_\mu$ by
\begin{eqnarray}
\label{covder}
D^+_\mu\Phi_x&=&U_{x,\mu}\Phi_{x+\mu}U^\dagger_{x,\mu}-\Phi_x\,,\\
D^-_\mu\Phi_x&=&\Phi_x-U^\dagger_{x-\mu,\mu}\Phi_{x-\mu}
U_{x-\mu,\mu}\,.\nonumber
\end{eqnarray}
Before continuing, we note that both $\cw_{x,\mu}$ and
$D^-_\mu\cw_{x,\mu}=\cw_{x,\mu}-U^\dagger_{x-\mu,\mu}\cw_{x-\mu,\mu}
U_{x-\mu,\mu}$ live in the coset space $\cg/\ch$.
For $\cw_{x,\mu}$ this follows from
\begin{eqnarray}
\tr([U_{x,\mu}T^iU^\dagger_{x,\mu},T^i]T^j)=
\ \tr(U_{x,\mu}T^iU^\dagger_{x,\mu}[T^i,T^j])=0\,.
\nonumber
\end{eqnarray}
For the second term in $D^-_\mu\cw_{x,\mu}$, this follows from
\begin{eqnarray}
\tr(T^jU^\dagger_{x,\mu}[U_{x,\mu}T^iU^\dagger_{x,\mu},
T^i]U_{x,\mu})=\ \tr([U_{x,\mu}T^jU^\dagger_{x,\mu},
U_{x,\mu}T^iU^\dagger_{x,\mu}]T^i)=0\,.
\nonumber
\end{eqnarray}
The off-shell gauge-fixing lattice action can now be expressed as
\begin{eqnarray}
\cl_{gf}^L
&=&
2\,\tr\!\left( [T_i,D^+_\mu C_x]\,
      [U_{x,\m} T_i  U_{x,\m}^\dagger , D^+_\mu \Cbar_x] \right)
  - 2i\,\tr\!\left( b_x\, D^-_\mu \cw_{x,\m} \right)
\label{tutu2}
\NXT
  && - 2i\,\tr\!\left( \{ D^+_\mu C_x , \Cbar_x \}\, \cw_{x,\m} \right)
 + \xi g^2 \sbar s\ \tr\left(\Cbar_xC_x\right)\,,
\nonumber
\end{eqnarray}
where $\sbar s\ \tr(\Cbar_xC_x)$ is still given by Eq.~(\ref{lgfwcc}).

Integrating out the auxiliary field
we then find that on the lattice $\tr\cf^2/(\xi g^2)$
(\cfdot\ Eqs.~(\ref{gf}, \ref{lgfonshell}) is
replaced by
\begin{equation}
\label{lnonab}
\cl_{G/H}\equiv{1\over \xi g^2}\
\tr\left(D^-_\mu\cw_{x,\mu}\right)^2\,.
\end{equation}
It is simple to check that $D^-\cw_{x,\mu}$ transforms covariantly
under $H$.  It follows that in the classical continuum limit
$D^-_\mu\cw_{x,\mu}\to D_\mu(V)W_\mu(x)= \cd_\mu(A)W_\mu(x)$.

The ghost part of the lagrangian derived from Eq.~(\ref{lgflattice})
is
\begin{eqnarray}
\label{lghost}
\cl_{ghost}&=&
2\ \tr\left([T^i,D^+_\mu C_x][U_{x,\mu}T^iU^\dagger_{x,\mu},
D^+_\mu\Cbar_x]\right)\\
&&-i\ \tr\left(2\{D^+_\mu C_x,\Cbar_x\}\cw_{x,\mu}+
\{\Cbar_x,C_x\}D^-_\mu\cw_{x,\mu}\right)\nonumber\\
&&+\xi g^2\Biggl(\tr\left(\left(\Cbar^2\right)_{\cg/\ch}
\left(C^2\right)_{\cg/\ch}\right)
+{3\over 4}\ \tr\left(\{\Cbar,C\}_{\cg/\ch}\right)^2
-\tr\left(\Xt^2\right)\Biggr)\,.\nonumber
\end{eqnarray}
It is straightforward to verify that the ghost part of
Eq.~(\ref{lgfonshell}) is recovered in the classical continuum
limit, by replacing first $U_{x,\mu}T^iU^\dagger_{x,\mu}\to T^i$,
and then using the relation
\begin{equation}
\label{trti}
\sum_i\tr\left([T^i,A][T^i,B]\right)=-\ \tr\left(A_{\cg/\ch}B_{\cg/\ch}
\right)\,,
\end{equation}
which follows from relation (\ref{ffd}).

An alternative lattice gauge-fixing action is given by
\begin{equation}
\label{lfgl2}
{\cl'}^L_{gf}=-s\sbar\ \tr(\cw^2+\xi g^2\Cbar C)\,.
\end{equation}
This action has the same classical continuum limit, and both lattice
actions are invariant under flip symmetry (as can be seen
most easily from their definition as the eBRST and anti-eBRST
variations of a flip-invariant pre-potential).

There are several further remarks we wish to make before
concluding this section.  First, it is straightforward to
check that, in both cases, the free kinetic term for the ghost
fields contains a nearest-neighbor discretization of the
laplacian.  This implies that no species doubling occurs in the
ghost sector on the lattice.

Second, both lattice actions are still invariant under the discrete
$SU(N)$ subgroup $\St_N$ introduced in the previous
section.  Under $\St_N$, the lattice gauge field transforms as
\begin{equation}
\label{snu}
U_\mu\to\Pt^k_{(AB)}U_\mu\left(\Pt^k_{(AB)}\right)^\dagger\,,
\end{equation}
consistent with Eq.~(\ref{snadj}).  The invariance of
Eq.~(\ref{lgflattice}) follows from Eq.~(\ref{hgen}).
Finally, the lattice actions are invariant under ghost-$SU(2)$.

\vspace{5ex}
\noindent {\large\bf 4.~Evading Neuberger's theorem}
\secteq{4}
\vspace{3ex}

It has been shown that a gauge-fixed Yang--Mills theory with
conventional BRST symmetry is not well-defined non-perturbatively
\cite{hnnogo}.  With ``non-perturbative" we refer to a lattice
definition of the theory which maintains exact BRST symmetry.
The theorem states that the partition function of such a theory,
as well as the (un-normalized) expectation
values of any gauge-invariant operator,
vanish identically.  What we wish to
demonstrate in this section is that equivariant gauge fixing
as described in the previous sections circumvents this problem
\cite{schaden}.

It is instructive to review the proof of the theorem first,
in order to see exactly what changes the conclusion in the
equivariant case; the key ingredient is the presence of
four-ghost terms in the gauge-fixing lagrangian~(\ref{lgfdetail}).

In the standard case, the lattice partition function can be written as
$Z_{BRST}(1)$ where
\begin{equation}
\label{brstz}
Z_{BRST}(t)
=\int[dU][db][dc][d\cbar]\ e^{-S_{inv}(U)-S_{gf}(t;U,c,\cbar,b)}\ .
\end{equation}
Here $S_{inv}$ is gauge invariant and depends only on
the physical fields.  $dU$ is the Haar measure on $G$, and with
the notation $[\ ]$ we indicate products over all sites and links
(for the gauge fields) or group indices (for ghost and auxiliary
fields).  $S_{inv}$ may include source terms
for gauge-invariant operators.  The gauge-fixing term
\begin{equation}
\label{gfbrst}
S_{gf}(t)= \sum_x \left(t\,\shat\ \tr(2\cbar\cf(U))+
\xi g^2\ \tr(b^2)\right)
\end{equation}
is a function of the standard ghost fields $c^a$ and $\cbar^a$,
with one pair for each generator of $G$, and an auxiliary
field $b^a$ for each generator as well.  Following ref.~\cite{hnnogo},
we introduced a parameter $t$ in front of the first term in
$S_{gf}$.  Standard BRST transformations are
\begin{eqnarray}
\label{brst}
\shat U_{x\mu}&=&i(U_{x,\mu}c_{x+\mu}-c_xU_{x,\mu})\,,\\
\shat c&=&-ic^2\,,\ \ \
\shat\cbar=-ib\,,\ \ \
\shat b=0\,,\nonumber
\end{eqnarray}
and invariance of the action follows immediately from the
fact that $\tr(b^2)=\shat\ \tr(i\cbar b)$ and
nilpotency of $\shat$, $\shat^2=0$.

The proof of Neuberger's theorem now follows very easily.
First one observes that
\begin{equation}
\label{dzdt}
  dZ_{BRST}/dt=-\svev{\shat\ \tr(2\cbar\cf(U))}_{\rm un-normalized}=0 \,,
\end{equation}
because of BRST invariance.
For $t=0$ the integral is well defined on the (finite-volume)
lattice because of the compactness of the gauge-fields,
and it follows that $Z(1)=Z(0)=0$.  The latter equality
follows immediately from the Grassmann integration rules
because the integrand for $Z(0)$ does not contain any ghosts.

Turning to the equivariant case, the path integral can be written as
$Z_{eBRST}(1)$ where
\begin{equation}
\label{ebrstz}
Z_{eBRST}(t)
=\int[dU][db][dC][d\Cbar]\ e^{-S_{inv}(U)-S^L_{gf}(t;U,C,\Cbar,b)}\,.
\end{equation}
Now there are only ghost, anti-ghost, and auxiliary
fields $C^\alpha$, $\Cbar^\alpha$ and $b^\alpha$ for the
coset generators $T^\alpha$ in $\cg/\ch$.  The gauge-fixing
part now corresponds to Eq.~(\ref{lgflattice}), and can be written as
\begin{equation}
\label{sgf}
S^L_{gf}(t)=\sum_x \left(t\, s\ \tr(2\Cbar\cf(U))
-\xi g^2\, s\sbar\ \tr(\Cbar C)\right)\,.
\end{equation}
Again, we have that $dZ_{eBRST}/dt=0$, because of eBRST invariance.
But now $Z_{eBRST}(0)\ne 0$,
due to the presence of ghost fields in the term proportional
to $\xi$ in $S^L_{gf}$.  Note that we {\em cannot} take
$\xi\to 0$, because the $b$ integrals do not converge in that
limit.  For $t=0$, we find that
\begin{eqnarray}
\label{zatzero}
Z_{eBRST}(1)&=&Z_{eBRST}(0)\\
&=&\int[dU]\ e^{-S_{inv}(U)}
\int[db][dC][d\Cbar]\
\exp\left[{\xi g^2\sum_x\ s\sbar\ \tr\left(\Cbar C \right)}\right]\,.
\nonumber
\end{eqnarray}
For $t=0$ the ghosts are decoupled from the lattice gauge field and,
moreover, the ghosts' partition function factorizes as the product
of independent single-site integrals. The single-site ghost integral
in this expression can be simplified further.
Going back to Eq.~(\ref{lgfwcc}), this integral may be written as
$Z_{ghost}(\xi g^2,1)$ where
\begin{equation}
\label{zghost}
Z_{ghost}(\xi g^2,t')=\int db\, dC\, d\Cbar\
\exp\left[{i\xi g^2\, \sbar\ \tr\left(bC-t'\Cbar C^2\right)}\right]\,,
\end{equation}
where we have introduced another parameter $t'$.  Using anti-eBRST
invariance, we see that
$\partial Z_{ghost}/\partial t'
=\partial Z_{ghost}/\partial \xi=0$,
and thus that
\begin{eqnarray}
\label{finalc}
Z_{ghost}(\xi g^2,1)&=&Z_{ghost}(1,0)\\
&=&\int db\, dC\, d\Cbar\ \exp\left[{
i\sbar\ \tr\left(bC\right)}\right]\nonumber\\
&=&\int db\, dC\, d\Cbar\ \exp\left[{
-\tr\left(b^2-\Xt^2\right)}\right]\,.\nonumber
\end{eqnarray}
Using this result, the rest of the proof that $Z_{ghost}(1,0)>0$ is technical,
and is relegated to Appendix A.

We see that in the equivariant case, the
ghost-field integral does not vanish, and thus the full path integral
does not vanish.  The underlying reason for the difference
with the standard case is that $\tr(b^2)$ itself cannot be written
as the eBRST variation of anything, because $s\ \tr(b^2)\ne 0$.
In order to build an eBRST invariant action, four-ghost terms
are needed, and these render the ghost integral non-zero.
It follows that the equivariantly gauge-fixed
partition function, $Z_{eBRST}(1)$, is equal---up to a
non-zero multiplicative constant $(Z_{ghost}(1,0))^V$
where $V$ is the volume in lattice units---to the partition function without
gauge fixing, which gives the standard lattice definition
of a Yang--Mills theory.  We recall that $S_{inv}$ may
contain source terms for any gauge-invariant operator
constructed out of the link variables $U_{x,\mu}$.  Therefore,
we see that the equivariantly gauge-fixed partition
function generates gauge-invariant correlation functions
which are rigorously equal to those generated by the
non-gauge-fixed theory, for any finite volume and any finite lattice spacing.

It is instructive to restate this result in a somewhat
different way.  Performing a gauge transformation on
$U_{x,\mu}$,
\begin{equation}
\label{ugt}
U^\phi_{x,\mu}=\phi_xU_{x,\mu}\phi^\dagger_{x+\mu}\,,
\end{equation}
and multiplying the partition function by
\begin{equation}
\label{fptrick}
1=\int [d\phi]\,,
\end{equation}
(again $d\phi$ is the normalized Haar measure)
the partition function may be written as
\begin{eqnarray}
\label{zhiggsp}
Z_{eBRST}(t)&=&\int [dU]\ Z_{orbit}(t;U)\ e^{-S_{inv}}\,,\\
Z_{orbit}(t;U)&=&\int [d\phi][db][dC][d\Cbar]
\ e^{-S^L_{gf}(t;U^\phi,C,\Cbar,b)}\,.\nonumber
\end{eqnarray}
$S^L_{gf}$ is now invariant under the ``orbit" eBRST
transformations
\begin{eqnarray}
\label{orbitebrst}
s\phi&=&-iC\phi\,,\ \ \ \ \ sU=0\,,\\
sC&=&\left(-iC^2\right)_{\cg/\ch}\,,\ \ \ s\Cbar=-ib\,,
\ \ \ sb=[X,\Cbar]\,.\nonumber
\end{eqnarray}
Note that $s$ is the same eBRST transformation as before,
but the transformation of the gauge fields is now
``carried" by the $G$-valued field $\phi$.
(The anti-eBRST rules may again be obtained by applying a flip transformation.)
Eq.~(\ref{zatzero}) can now be restated by observing that
$Z_{orbit}$ is independent of the gauge
field.  This can again be seen by varying the parameter
$t$ (\cfdot\ Eq.~(\ref{sgf})).  It follows that
$dZ_{orbit}/dt=0$, and thus $Z_{orbit}(t;U)=Z_{orbit}(0;U)$,
the latter being independent of $U$.  In other words,
$Z_{orbit}(t;U)$ is a topological field theory.

One of the consequences is that the equivariantly gauge-fixed
theory is unitary if the non-gauge-fixed theory is.
This is a rigorous result, as all manipulations in this
section are valid for the lattice path integral in a finite volume.

\vspace{5ex}
\noindent {\large\bf 5.~Perturbative unitarity}
\secteq{5}
\vspace{3ex}

While the results obtained in the previous two sections are
interesting in their own right, our aim is to apply them
to the construction of lattice chiral gauge theories.
For this goal, the gauge group $G=SU(N)$ will need to be fixed completely
and non-perturbatively: a complete gauge-fixing action will
have to be included in the definition of the lattice theory.

Gauge fixing of the remaining subgroup $H$ is, of course, also
needed if one wishes to develop perturbation theory for
any (continuum or lattice) equivariantly gauge-fixed theory.
In the case at hand, the subgroup $H=U(1)^{N-1}$ is abelian,
and significant simplification occurs.
In order to fix an abelian invariance, only a simple
gauge-fixing term like $\cl_L=(1/2\alpha)\sum_i(\partial_\mu A^i_\mu)^2$
is necessary. There is no need to introduce any new ghost fields.
The addition of $\cl_L$ does break eBRST invariance, and this
raises the issue of unitarity.  Slavnov--Taylor
identities, derived from (e)BRST invariance, are a key
ingredient in the study of unitarity. Therefore one has to re-establish
unitarity in the presence of $\cl_L$.

\begin{figure}[t]
\vspace*{0.4cm}
\centerline{
\epsfysize=2.5cm
\put(-175,0){\epsfbox{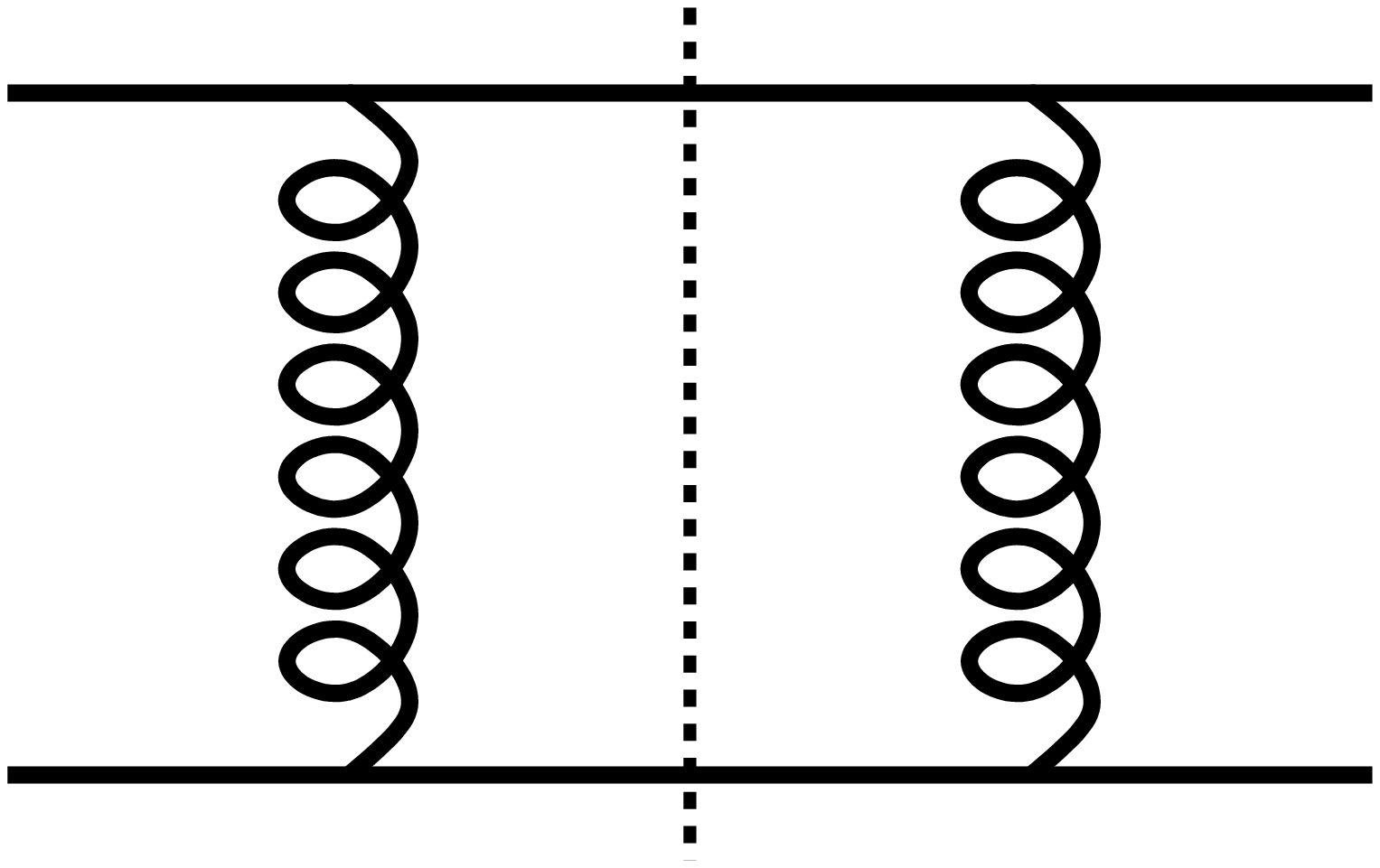}}
\epsfysize=2.5cm
\put(0,0){\epsfbox{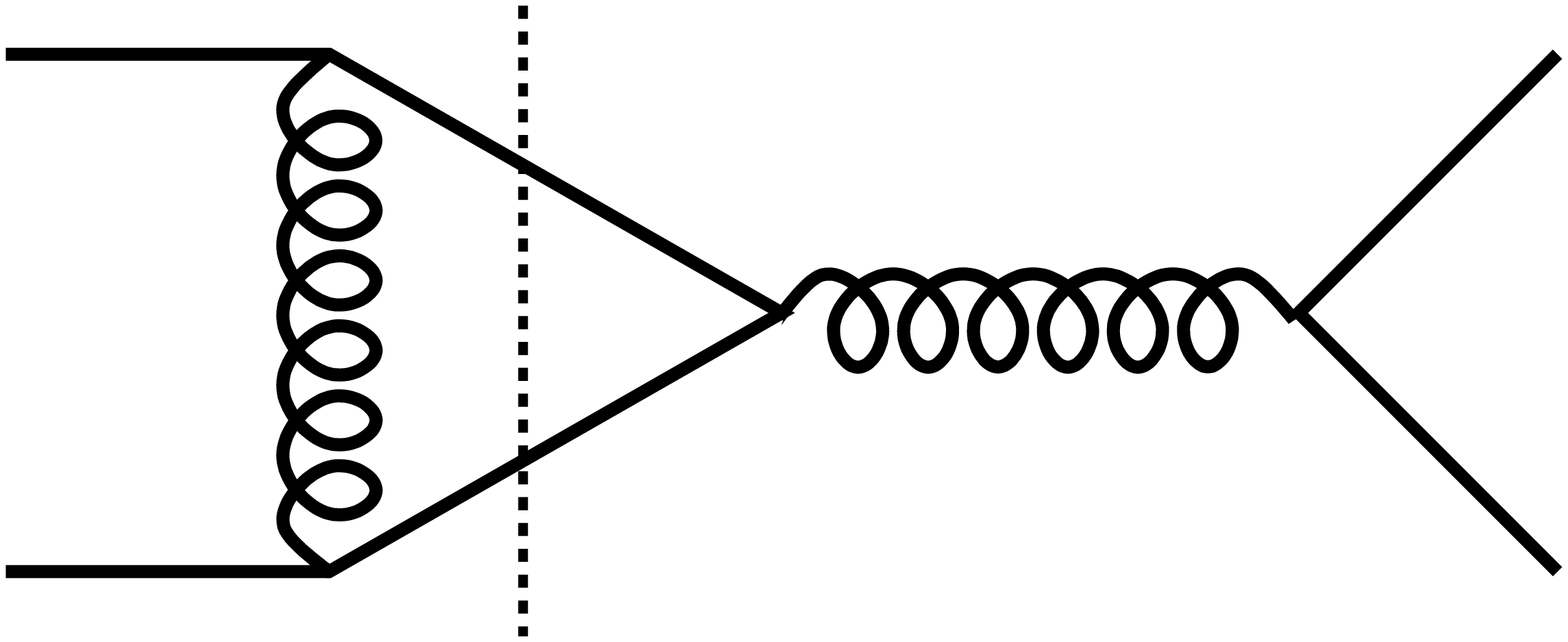}}
\epsfysize=2.5cm
\put(-175,-120){\epsfbox{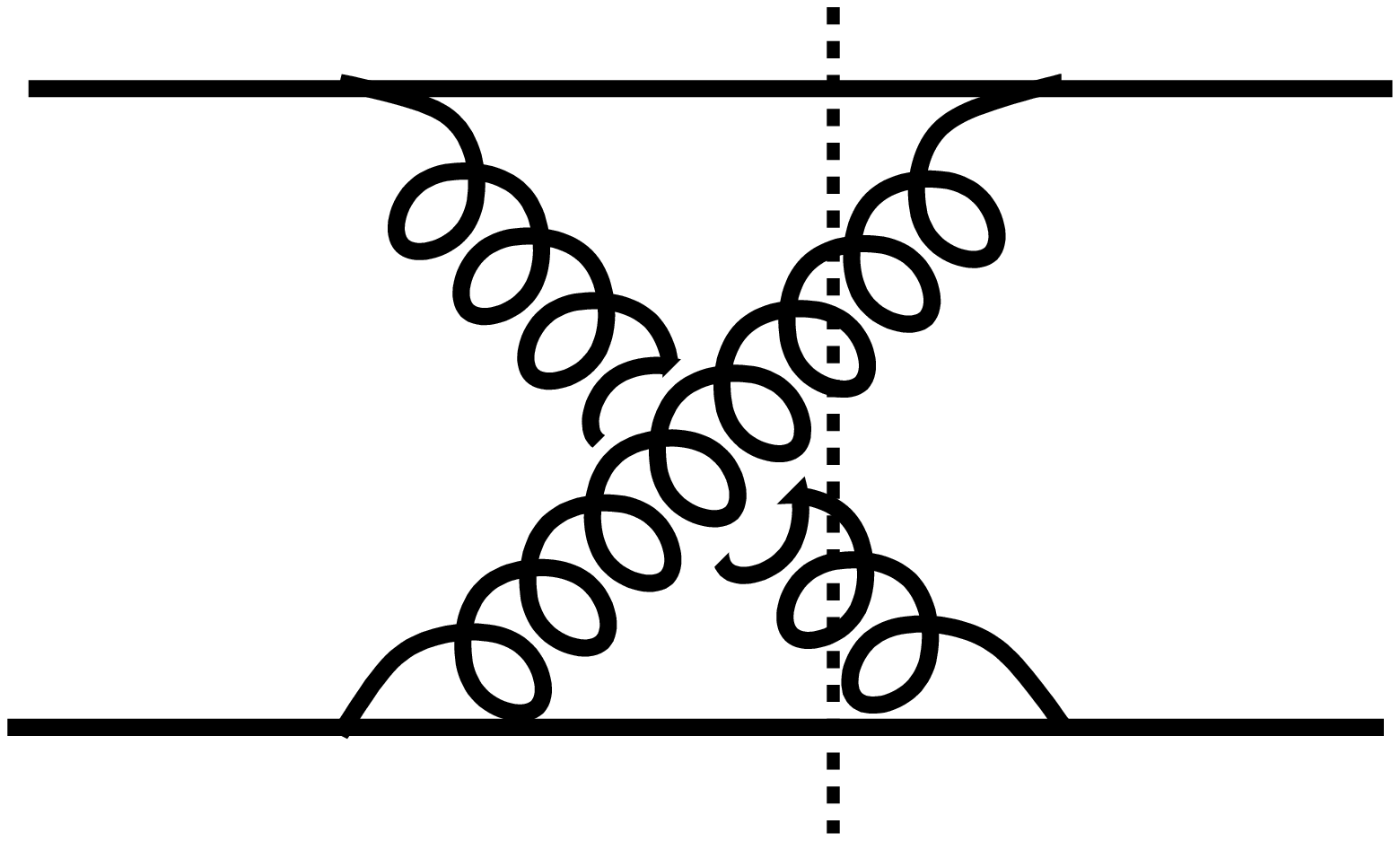}}
\epsfysize=2.5cm
\put(0,-120){\epsfbox{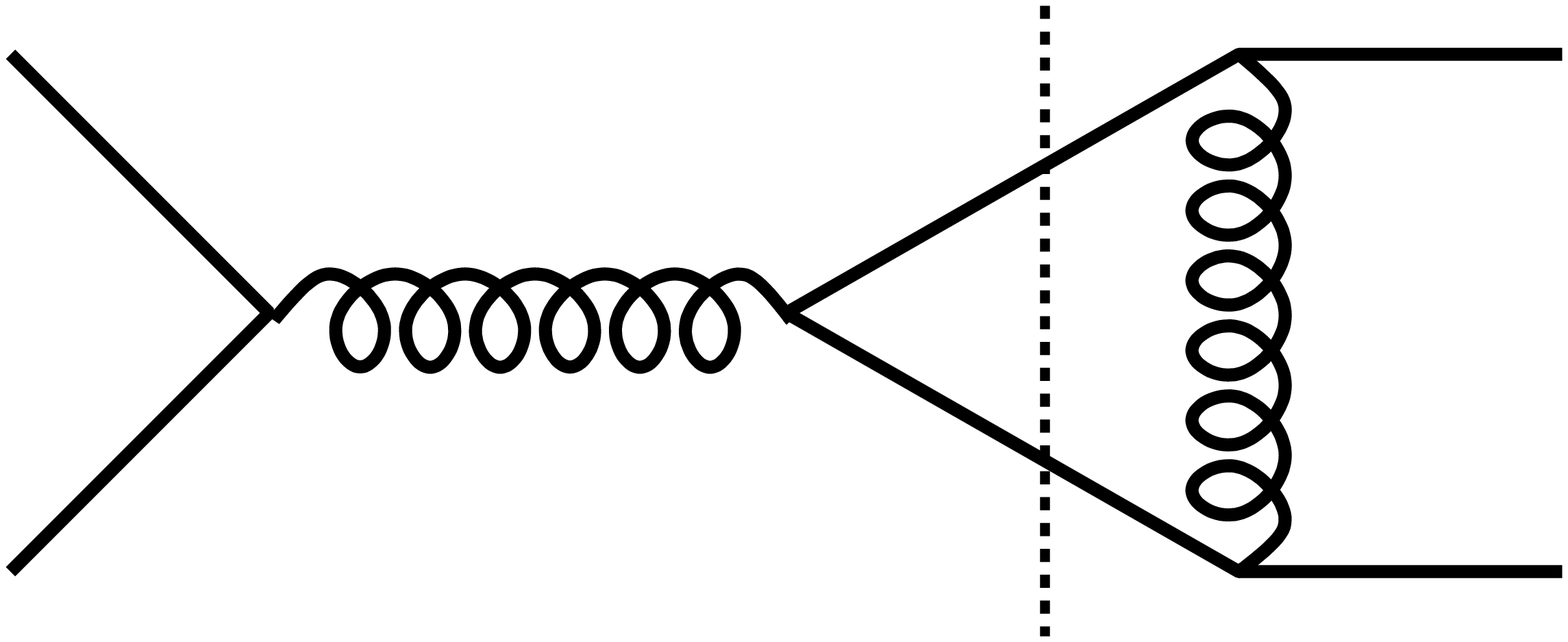}}
}
\vspace*{0.6cm}
\begin{quotation}
\caption{ \noindent {\it Cut diagram with two fermions in the intermediate
state. Solid lines denote fermions, curly lines denote gluons.}}
\end{quotation}
\end{figure}

Here we will address this question in (continuum) perturbation theory.
The conclusion is
that, as expected, $H$ gauge fixing does not spoil the unitarity of the
theory. Heuristically, this is easy to understand.
The eBRST version of Yang--Mills
theory is rigorously the same as the non-gauge-fixed
version, if one restricts oneself to the physical sector,
\ie\ to gauge-invariant correlation functions of operators
built only from the physical fields.  Gauge-fixing either
theory completely in order to develop perturbation theory
(which for the eBRST version implies only
fixing $H$) should not change this result.

In the context of perturbation theory,
it is convenient to introduce an $\ch$-ghost sector. As we will see,
in the continuum the new ghosts are free fields that
merely serve as a device to generate the relevant Slavnov--Taylor
identities. Since they should decouple anyway in the
continuum limit, no $\ch$-ghosts will be introduced in the actual
lattice construction of chiral gauge theories.
(The eBRST and $H$-BRST
identities of the target continuum theory are sufficient to determine
the lattice counter terms to all orders; see also ref.~\cite{morebrst}.)
As for the exactly eBRST-invariant
(and $H$ un-gauge fixed) lattice theory defined in Sect.~3,
the $H$-gauge-fixing  sector
is an extraneous analytic device used to set up perturbation theory,
in the same way as gauge fixing is needed to set up
perturbation theory for the standard, fully gauge invariant lattice
Yang-Mills theory.

\begin{figure}[t]
\vspace*{0.4cm}
\centerline{
\epsfysize=2.5cm
\put(-175,0){\epsfbox{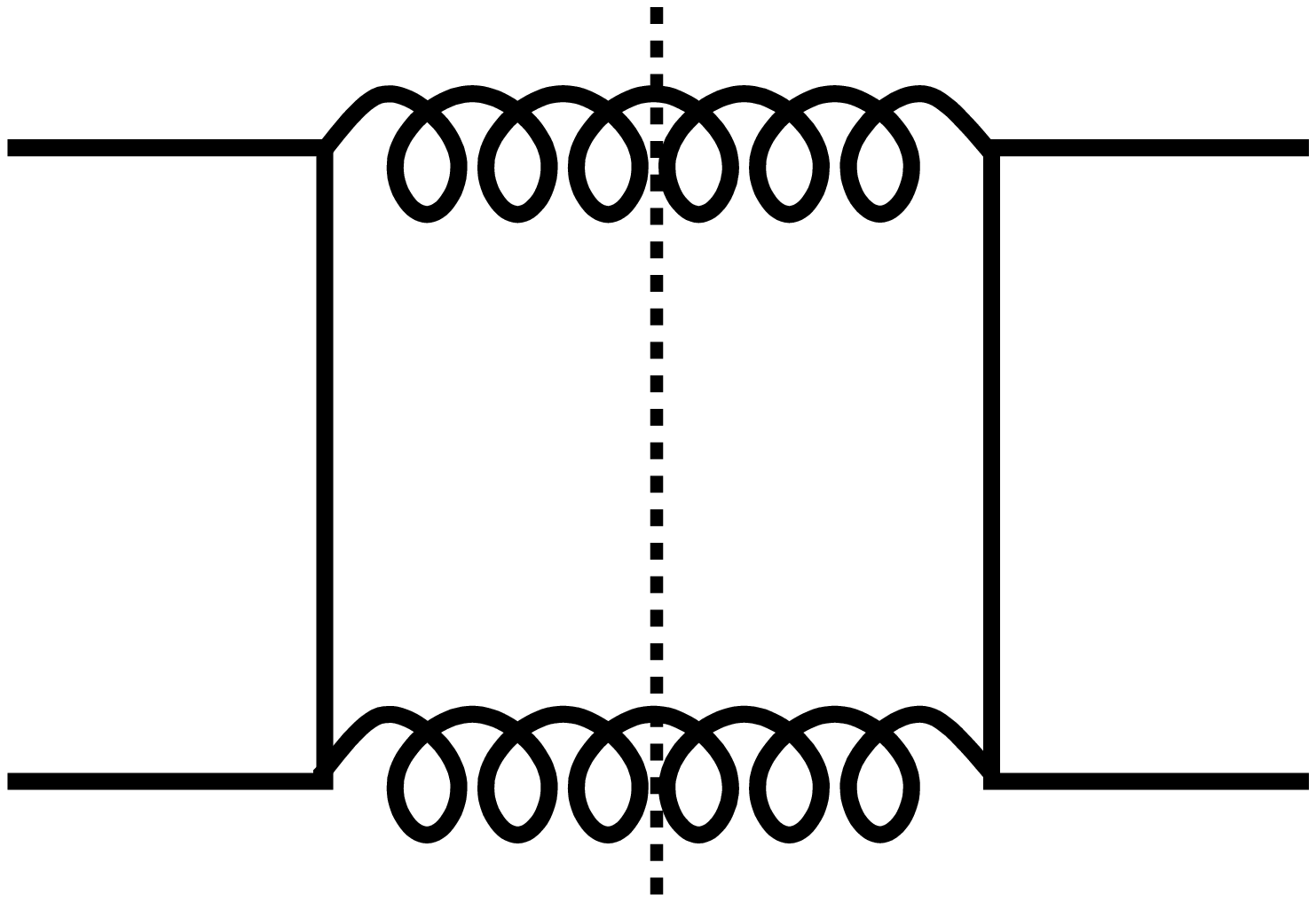}}
\epsfysize=2.5cm
\put(0,0){\epsfbox{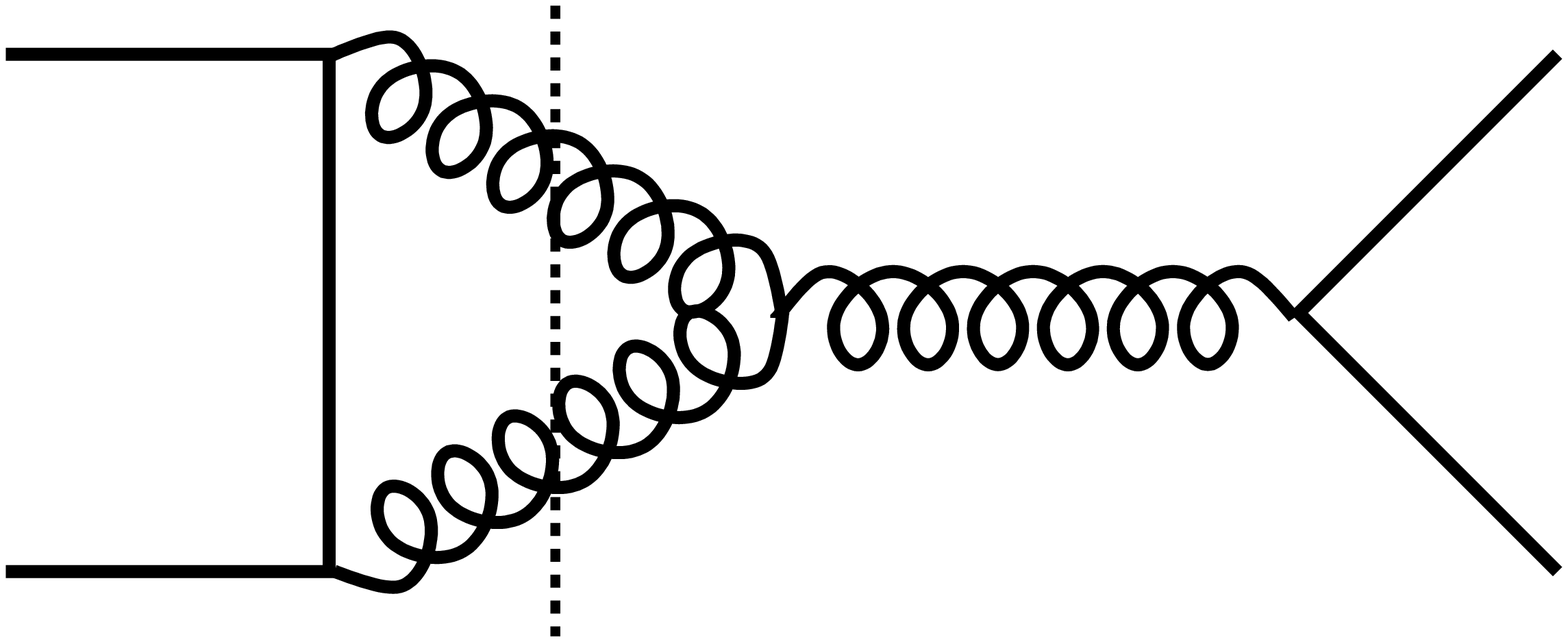}}
\epsfysize=2.5cm
\put(-175,-120){\epsfbox{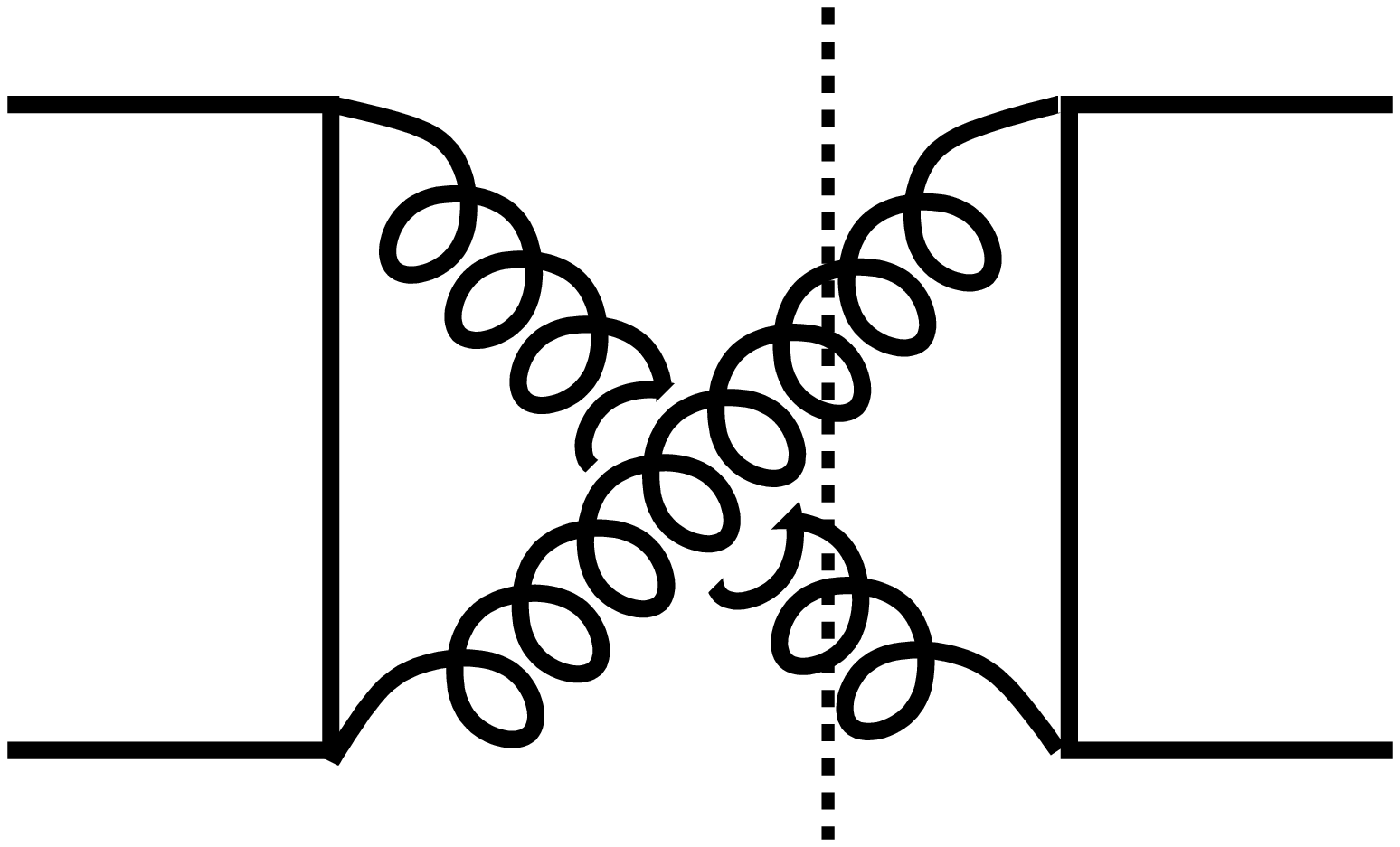}}
\epsfysize=2.5cm
\put(0,-120){\epsfbox{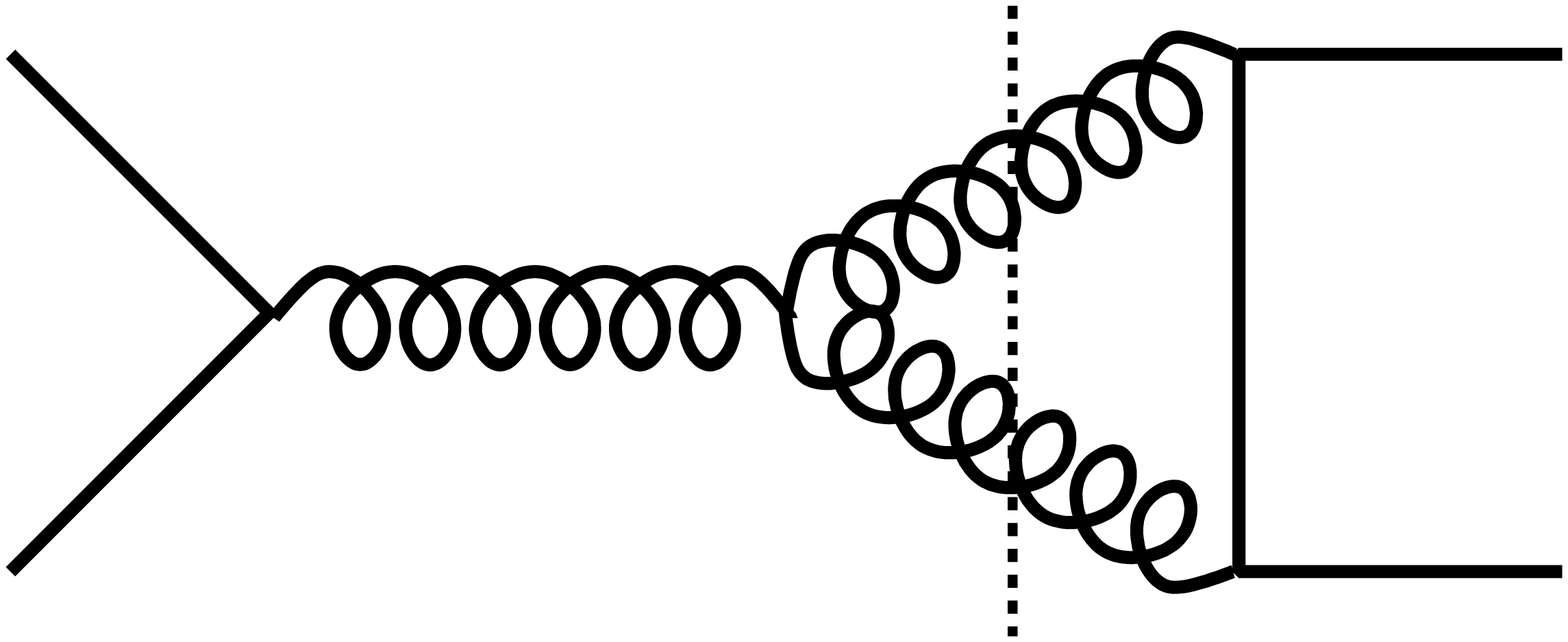}}
}
\vspace*{0.6cm}
\begin{quotation}
\caption{ \noindent {\it Cut diagrams with two gauge bosons in the
intermediate state.
}}
\end{quotation}
\end{figure}

Emphasizing again that this is only done for perturbative
investigations, we introduce $\ch$ ghost fields
$\chi^i$, $\chibar^i$ and auxiliary field $\beta^i$,
and define $\ch$-BRST transformation rules
\begin{eqnarray}
\label{hbrst}
s_H\Psi&=&\delta_\chi\Psi\,,\ \ \ \ \ s_HA^i_\mu=\partial_\mu\chi^i\,,\\
s_H\chi^i&=&0\,,\ \ \ \ \ s_H\chibar^i=-i\beta^i\,,\ \ \ \ \
s_H\beta^i=0\,,
\nonumber
\end{eqnarray}
where $\Psi$ stands for any of the fields $V_\mu$ (or $U_\mu$),
$C$, $\Cbar$ and $b$. Under eBRST transformations, the new fields
$\chi^i$, $\chibar^i$ and $\beta^i$ are invariant by definition.
With
\begin{equation}
\label{preH}
  \ck = \int d^4x \left(
  \chibar^i(\partial_\mu A^i_\mu)+{i\over 2}\alpha g^2
  \chibar^i\beta^i\right)\,,
\end{equation}
the additional (off-shell) gauge-fixing action
can now be written as
\begin{equation}
\label{ll}
  S_L = s_H\, \ck
  = \int d^4x \left( -i\b^i \partial_\mu A^i_\mu - \chibar \bo \chi
    + {\alpha g^2\over 2} \beta^2 \right)\,,
\end{equation}
where $\alpha$ is a gauge-fixing parameter not necessarily
equal to $\xi$ (\cfdot\ Eqs.~(\ref{lgfpartly}, \ref{lgf})).
A consequence of having $s$ vanish when acting on any
$\ch$-ghost-sector field is that \cite{schaden}
\begin{equation}
\label{ghalgebra}
\{s,s_H\}=0\,.
\end{equation}
This makes it possible to ``port" eBRST
identities to the fully gauge-fixed theory.
In the equivariantly but not $H$-gauge-fixed theory,
the eBRST identities of interest are of the form
\begin{equation}
\label{ebrstid}
\langle s\co\rangle=0\,,
\end{equation}
with $\co$ any operator invariant under $H$ gauge symmetry,
\ie\ operators for which $s_H\co=0$.
In the fully gauge-fixed theory, in which $S_L$ is added,
we have that
\begin{equation}
\label{ebrstidagain}
\svev{s\co}
=-\svev{\co\ ss_H\ck}
=\svev{\co\ s_Hs\,\ck}=0\,.
\end{equation}
The last step follows
because $\co$ is invariant under $s_H$.\footnote{We assumed that the
operator $\co$ is anti-commuting; if it is commuting, $s\co$ is
anti-commuting, and thus $\langle s\co\rangle=0$ trivially.}
This proves that the same eBRST identities hold in the fully
gauge-fixed theory as well.  In what follows below we will also
have use for the case that $\co$ is not invariant under $H$,
in which case we obtain
\begin{equation}
\label{ononinv}
\svev{s\co}=
\svev{(s_H\co) (s\,\ck)}
= -\svev{(s_H\co)\
\int d^4x\, f_{i\alpha\beta}(\partial_\mu\chibar^i)W^\alpha_\mu C^\beta}\,.
\end{equation}
Of course, in the fully
gauge-fixed theory one also has Slavnov--Taylor identities
derived from $s_H$, and they play a role in proving
unitarity as well.

A comment on the appearance of $\chibar^i$ in the latter
identity is in order.
First, if $s_H\co$ does not contain the field $\chi^i$, $\svev{s\co}$
vanishes identically.  Otherwise, we may carry out the contractions
and replace $\langle\chi^i\chibar^j\rangle$ by its (tree-level)
propagator, because $\chi^i$ and $\chibar^i$ are free fields.
Indeed, we need not have introduced the $H$ ghosts into the theory;
they are just a convenient vehicle for deriving the desired
Slavnov--Taylor identities. In the remainder of this section
we will keep $\chi^i$ and $\chibar^i$
as a ``book-keeping device." They will not be part of our definition
of lattice chiral gauge theories in the next section.

As a first application we will prove that all scattering
amplitudes are independent of the gauge-fixing
parameters $\xi$ and $\alpha$ to all orders in perturbation theory.
To do this, it is convenient to
use the formalism in which the auxiliary fields $b$ (\cfdot\
Eq.~(\ref{lgfdetail})) and $\beta$ (\cfdot\ Eq.~(\ref{ll})) are kept.
We show in Appendix B that Feynman rules can be consistently
formulated in this framework.  Consider the $\xi$ dependence of the
expectation value of a (commuting) operator $\co$,
\begin{eqnarray}
\label{ddxi}
{1\over g^2}{d\over{d\xi}}\svev{\co}
&\!\!=&\svev{\co\int d^4x\, s\sbar(\Cbar C)} \\
&\!\!=&\!-\svev{ (s\co) \int d^4x\,\sbar(\Cbar C)}
+\svev{\co (ss_H \ck) \int d^4x\,\sbar(\Cbar C)}
\nonumber\\
&\!\!=&\!-\svev{ (s\co) \int d^4x\,\sbar(\Cbar C)}
+\svev{ (s_H\co) (s\,\ck) \int d^4x\,\sbar(\Cbar C)}\,, \nonumber
\end{eqnarray}
where the last equality follows from Eq.~(\ref{ghalgebra}) and
the fact that $\Cbar C$ is $H$-invariant.  Now if we take $\co$
to be the product of matter fields, such as quark and anti-quark fields
$q$ and $\qbar$, we have that both $sq$ and $s_H\, q$,
as well as $s\qbar$ and $s_H\,\qbar$, are
non-linear in these fields. Therefore, the right-hand side of
Eq.~(\ref{ddxi}) vanishes if we analytically continue to Minkowski space,
amputate the fermion legs and
put them on shell. In Appendix C we generalize the argument to scattering
amplitudes containing gauge bosons on the external legs.
We conclude that all scattering amplitudes are
independent of $\xi$, as should be the case.  A similar, even
simpler, argument shows $\alpha$ independence as well.

We comment in passing that invariance under a continuous change
of parameters in the gauge-fixing action has played an important role
in the preceding section too (see \eg Eq.~(\ref{zatzero})).
The conclusions of the previous section
are stronger because the non-perturbative setup allows us to set
to zero many terms in the gauge-fixing action,
while maintaining the extended eBRST invariance of the theory.

\begin{figure}[t]
\vspace*{0.4cm}
\centerline{
\epsfysize=2.5cm
\put(-140,120){\epsfbox{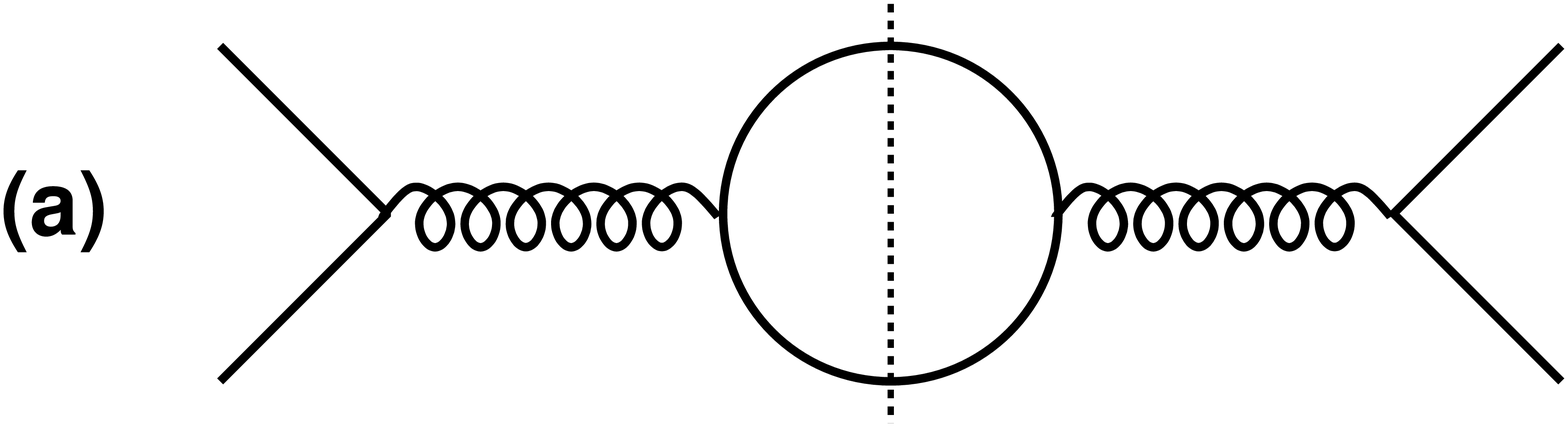}}
\epsfysize=2.75cm
\put(-140,0){\epsfbox{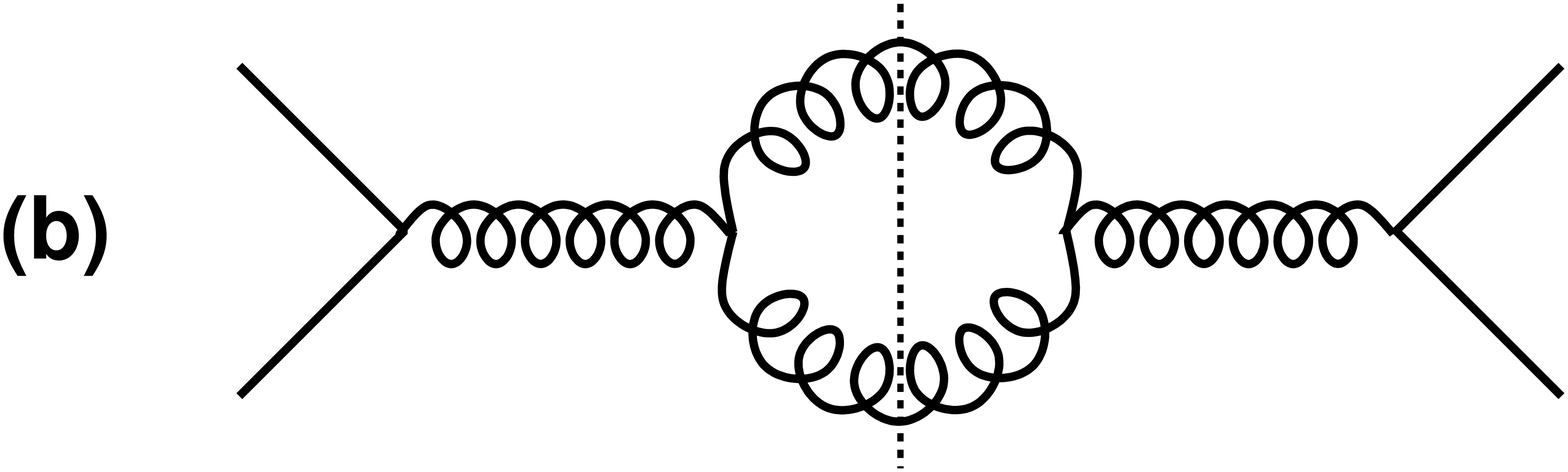}}
\epsfysize=2.5cm
\put(-140,-120){\epsfbox{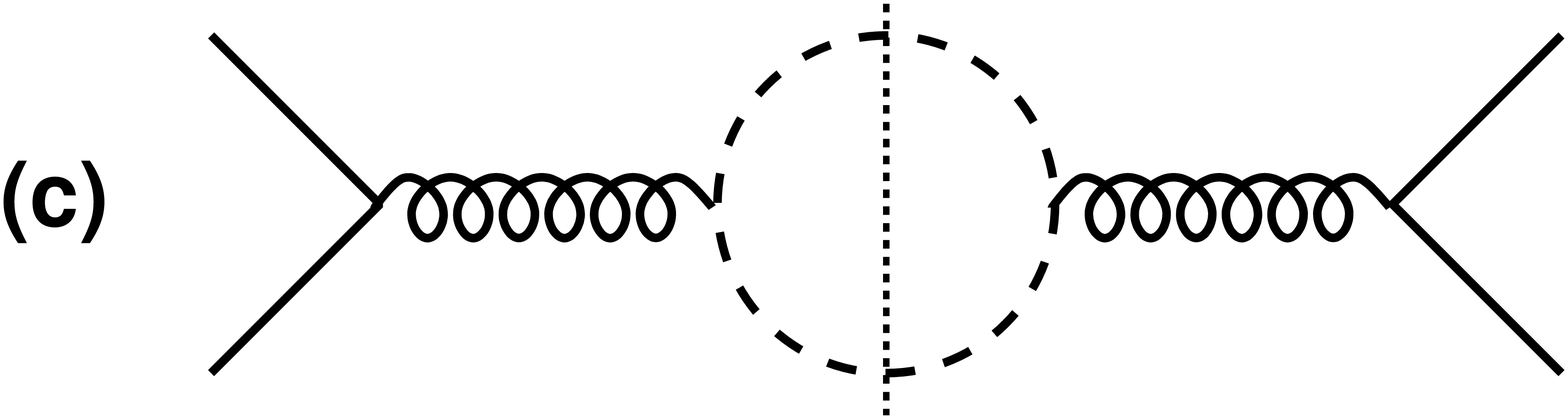}}
}
\vspace*{0.6cm}
\begin{quotation}
\caption{ \noindent {\it Cut diagrams containing a vacuum polarization
subdiagram. Dashed lines denote ghosts.
}}
\end{quotation}
\end{figure}

In order to make the discussion less abstract, we will now
work out an example that shows in more detail how
eBRST and $H$-BRST identities play a role in proving unitarity
in continuum perturbation theory.\footnote{
  We defer the discussion of the lattice and of chiral fermions
to the next section.}
With the above result in hand, we choose to work in the Feynman gauge,
\ie\ we now set $\xi=\alpha=1$,
thus simplifying our calculations. Our aim is to
demonstrate how unitarity works {\it vis-\`a-vis} the
unconventional gauge-fixing procedure introduced in this paper.
Thus, we will limit ourselves to the difference
between the eBRST case and the standard case.

We will consider a two-flavor vector gauge theory with
$G=SU(2)$, $H=U(1) \subset G$.
The gauge-sector lagrangian will be the sum of Eqs.~(\ref{YM}),
(\ref{lgfdetail}) and (\ref{ll}).  We add to this an
$SU(2)$ doublet of massless quarks, $q=(u,d)$.
For $SU(2)/U(1)$ one has $f_{\a\b\g}=0$, and
Eq.~(\ref{lgfdetail}) simplifies to
\begin{eqnarray}
\label{lexample}
\cl_{gf}&=&-2\ \tr\left(\Cbar\cd_\mu(A)\cd_\mu(A)C\right)
+2\ \tr\left([W_\mu,\Cbar][W_\mu,C]\right)\\
&&-2i\ \tr\left(b\cd_\mu(A)W_\mu\right)
+\xi g^2\left(\tr(b^2)-\tr\left(\Xt^2\right)\right)\,.
\nonumber
\end{eqnarray}
As usual, we will replace $A\to gA$ and $W\to gW$ in order to develop
the perturbation expansion in $g$.

The main example we wish to discuss is that of $d\ubar\to d\ubar$
scattering at one loop, in order to demonstrate that this amplitude
satisfies the optical theorem.\footnote{
  For a textbook discussion in standard Feynman gauge,
  see for instance ref.~\cite{peskincheng}.
}
There are three types of contribution
at one loop, shown in Figs.~1, 2 and 3.  Fig.~1 contains diagrams
with two-quark intermediate states, Fig.~2 contains diagrams with
two-gauge-boson intermediate states, and Fig.~3 contains diagrams with
one-loop vacuum polarization corrections to the one-gauge-boson
intermediate state.

We will only be interested in contributions of the type of Figs.~2 and
3(b,c), with only gauge boson and ghost contributions to the vacuum
polarization.  The diagrams of Fig.~1 and fermionic contributions to the
vacuum polarization (Fig.~3(a)) satisfy unitarity constraints separately, and
work just like in the standard case.  In the case of
$u\dbar$ scattering, the two-gauge-boson intermediate state is an $A$--$W^-$
intermediate state, with $A_\mu=V^3_\mu$ and $W^\mp_\mu=(V^1_\mu\pm
iV^2_\mu)/ \sqrt{2}$ in terms of the original $SU(2)$ gauge fields.
Note that $A$ is neutral under $H=U(1)$, while $W^\pm$ are charged.
Since there are no (interacting) neutral ghosts, it follows immediately that
there is no ghost-loop contribution to the vacuum polarization of $W^\pm$,
which means that Fig.~3(c) is absent for $d\ubar\to d\ubar$.
Therefore, when we cut the diagrams (as indicated in the figures),
contributions from unphysical polarizations of the gauge bosons will
have to cancel by themselves, without ``help" from any ghost-loop
diagrams.  This is where our example differs from the standard
case of a linear covariant gauge, in which of course ghosts
corresponding to every component of the gauge field are present.
In our case,
the gauge fixing takes place in two stages, of which the latter
is an abelian linear gauge fixing, for which no ghosts are needed.%
\footnote{Or, equivalently in perturbation theory, the corresponding
ghosts are free, as manifest in Eq.~(\ref{ll}).}

The optical theorem relates the cut diagrams of Figs.~2 and~3(b)
to the process $u\dbar\to AW^-$
(with both $A$ and $W^-$ transversely polarized).
If we consider only the contributions from physical polarizations of
intermediate $A$--$W^-$ states to the cut diagrams,
the optical theorem is evidently satisfied.
Thus, our aim here is to show how the eBRST identities of Eq.~(\ref{ebrstidagain})
plus the BRST identities derived from $s_H$ guarantee
that the contributions from all
unphysical polarizations to the cut diagram vanish.

The relevant eBRST identity is Eq.~(\ref{ononinv}) with
$\co=\Cbar^-(x)A_\nu(y)u(v)\dbar(w)$. Since we are interested
in the LSZ-amputated correlation function, the only terms which
contribute in Eq.~(\ref{ononinv})
are those from the linear terms in the eBRST and $H$-BRST transformation
rules. Analytically continuing to Minkowski space,
we thus obtain the identity (in Feynman gauge)
\begin{eqnarray}
\label{wlongid}
\svev{\partial^\mu W^-_\mu(x)A_\nu(y)u(v)\dbar(w)}_{os}&=&\\
&\hspace{-4cm}=&\hspace{-2cm}
\int d^4z\ \svev{\Cbar^-(x)\partial_\nu\chi(y)u(v)\dbar(w)\,
\partial^\mu\chibar(z)W^-_\mu(z)C^+(z)}_{os}\nonumber\\
&\hspace{-4cm}=&\hspace{-2cm}
\int d^4z\ \svev{\partial_\nu\chi(y)\partial^\mu\chibar(z)}
\svev{\Cbar^-(x)u(v)\dbar(w)W^-_\mu(z)C^+(z)}_{os}\,,
\nonumber
\end{eqnarray}
where the subscript $os$ means that all external lines have been
put on shell.
In the last step we made use of the fact that $\chi$ and $\chibar$
are free.  Fourier transforming on $x$ and $y$, this can be rewritten as
\begin{eqnarray}
\label{wlongidft}
k^\mu\svev{ W^-_\mu(k)A_\nu(\ell)u(v)\dbar(w)}_{os}&=&\\
&\hspace{-4cm}=&\hspace{-2cm}
\ell_\nu\int d^4z\ \svev{\chi(\ell)\partial^\mu\chibar(z)}
\svev{\Cbar^-(k)u(v)\dbar(w)W^-_\mu(z)C^+(z)}_{os}\,,
\nonumber
\end{eqnarray}
We will also need an $s_H$ identity for the longitudinal
part of $A$.  For this we take
$\co'=W^-_\mu(x)\chibar(y)u(v)\dbar(w)$.
{}From $\svev{ s_H\co'}=0$ we obtain
\begin{equation}
\label{alongid}
\ell^\nu\svev{ W^-_\mu(k) A_\nu(\ell)u(v)\dbar(w)}=0\,.
\end{equation}

Cutting the internal $W$ line
replaces the pole $1/(k^2+i\e)\to\delta(k^2)$ and leaves a tensor
$g_{\mu\rho}$ which may be expressed as a polarization sum
\begin{equation}
\label{poltensor}
g_{\mu\rho}
=-\sum_{n=1,2} \e^{(n)}_\mu \e^{(n)}_\rho
+\e^{+}_\mu\e^-_\rho+\e^{-}_\mu\e^+_\rho\,
\end{equation}
(similar statements apply to the cut $A$ line).
For a physical momentum
$k_\mu=(\kvec,|{\vec k}|)$,\footnote{
  In Minkowski space we use the metric
$g_{\mu\nu}={\rm diag}(-1,-1,-1,1)$.
}
the forward (or longitudinal)
and backward polarization vectors are defined as
\begin{equation}
\label{polvectors}
\e^{+}_\mu(\kvec)={1\over\sqrt{2}}(\khat,\ 1)={1\over{\sqrt{2}
|{\vec k}|}}\,k_\mu\,,\ \ \ \ \
\e^{-}_\mu(\kvec)={1\over\sqrt{2}}(-\khat,\ 1)\,,
\end{equation}
where $\khat$ is the unit vector in the direction of $\kvec$.
The normalized transverse polarization vectors $\e^{(n)}$ have
$\e^{(n)}_4=0$, and are orthogonal to $\e^{\pm}$.

We are now ready to prove unitarity of our example.  Cutting the
$W$ and $A$ lines leaves us with a sum over products of
amplitudes for $d\ubar\to W^-A$ scattering (in this case we will
refer to the $W^-$ and $A$ as ``out-going") and $W^-A\to d\ubar$
scattering (``in-going" $W^-$ and $A$).
First, Eq.~(\ref{alongid}) implies that any contribution
involving a forward polarized $A$ vanishes.
According to the polarization sum~(\ref{poltensor}),
this leaves us only with contributions for which the $A$ is transverse.
In this case, the $W^-$ can still be unphysical.
Again from Eq.~(\ref{poltensor}), either the in-going or the out-going
$W^-$ has to be forward polarized, and we may apply Eq.~(\ref{wlongidft}).
This equation tells us that the only non-vanishing contribution
with a forward-polarized $W^-$ has a backward polarized $A$
on the same side of the cut
(because $\e^\nu\ell_\nu = 0$ for the other three $A$ polarizations).
But this means that the $A$ on the other side
of the cut has a forward polarization, and we already showed
that all contributions involving a forward polarized $A$ vanish.
We conclude that none
of the unphysical polarizations contribute to the imaginary part
of the one-loop $d\ubar\to d\ubar$ amplitude.  We verified this
by explicit calculation.  In an explicit calculation, one makes
of course use of the fact that the $d$ and $\ubar$ external legs,
as well as the cut lines, are on shell.

A similar example can be worked out for the case that the in-going
quark and anti-quark have the same flavor.  In this case, the
two-gauge-boson intermediate state of interest is a $W^-W^+$ state.
One finds that now there is a $\cg/\ch$-ghost contribution to the vacuum
polarization, which is needed to cancel all contributions to the
cut diagram from unphysical $W$ polarizations.  The relevant
identities ensuring this cancellation are (again, in Feynman gauge)
\begin{equation}
\label{wwids}
\svev{\partial^\mu W^\pm_\mu(x)W^\mp_\nu(y)u(v)\ubar(w)}_{os}
=\svev{\Cbar^\pm(x)\partial_\nu C^\mp(y)u(v)\ubar(w)}_{os}\,.
\nonumber
\end{equation}
This is as one
would expect, since the $W$s belong to the ``non-abelian part"
of $G$.  We see that the way unitarity is enforced through the
(e)BRST identities is a ``combination" of how it works in the
standard abelian and non-abelian cases.  Again, we verified this
example by explicit calculation.

We close this section with a few comments on the differences
of our gauge-fixing and that of the standard (non-abelian)
Lorenz gauge.  The fact that our $G/H$ gauge-fixing condition $\cf(V)$
(\cfdot\ Eq.~(\ref{gf})) had to be $H$-covariant leads to extra
vertices in our gauge-fixed theory.  And, one finds indeed that
these additional vertices play a role in the explicit calculations
verifying unitarity in the examples discussed above.

The four-ghost vertices are another new type of vertices that appear
in our theory (\cfdot\ Eqs.~(\ref{lgfonshell}, \ref{lexample})).
They do not play a role in our examples
of one-loop unitarity -- one would have to go to higher loops
to encounter them.  However, there is a simple one-loop calculation
in which these new ghost vertices do play a role.  First note
that the ghosts have to be massless for (perturbative) unitarity
to work out.  In standard Lorenz gauge it is very easy to
see that indeed the ghosts have to be massless, because there is
a shift symmetry on the anti-ghost field in that case.
Clearly, there is no shift symmetry in our case.  Still, the
algebraic structure does not allow for a ghost mass term,
since it cannot be obtained from the eBRST transformation
of any operator.  But eBRST does imply the occurrence of the
four-ghost vertices, and one may indeed verify that
these vertices are needed so that no mass term is generated
from the one-loop ghost self energy.  This can be explicitly
checked in the lattice version of our theory.\footnote{
  In dimensional regularization, the massless one-loop tadpole is
  zero by definition.
}

\vspace{5ex}
\noindent {\large\bf 6.~Lattice chiral gauge theories}
\secteq{6}
\vspace{3ex}

We now come to the construction of chiral gauge theories on the
lattice in the framework of a fully gauge-fixed Yang--Mills theory
as developed in the preceding sections.  There are three steps
to this task.  First, we will specify the fermionic part
of the lattice action.  Then, after adding the chiral fermions
to the theory, we choose a lattice gauge-fixing action for the
$H$ gauge fixing, and we revisit the $G/H$ gauge fixing
on the lattice, for reasons to be discussed below.  The resulting
action for a chiral gauge theory is not exactly invariant under
(e)BRST on the lattice, and we will need to add counter terms;
this constitutes the final step.

As already explained in the introduction, the existence of a
systematic perturbative expansion in the coupling constant
is crucial in order to recover the target chiral gauge theory
in the continuum limit. We explain below how this is achieved.

We take our fermions to be left-handed and in some anomaly-free
(in general, reducible)
representation of the group $SU(N)$.\footnote{For $N=2$ the
fermion representation should be free of the Witten anomaly as
well.  Simple examples are a theory with two Weyl doublets (which is
not really chiral), or a theory with one Weyl fermion in the $3/2$
representation.}  In order to avoid doublers, we introduce
right-handed spectators \cite{smitswift} and add a Wilson
term, leading to
\begin{eqnarray}
\label{lfermion}
\cl_{fermion}&=&{1\over 2}\sum_\mu\left(
\psibar_{Lx}\gamma_\mu U_{x,\mu}\psi_{Lx+\mu}
-\psibar_{Lx+\mu}\gamma_\mu U^\dagger_{x,\mu}\psi_{Lx}\right)\\
&&+{1\over 2}\sum_\mu\left(
\psibar_{Rx}\gamma_\mu\psi_{Rx+\mu}
-\psibar_{Rx+\mu}\gamma_\mu\psi_{Rx}\right)\nonumber\\
&&-{r\over 2}\sum_\mu\left(
\psibar_{x}\psi_{x+\mu} +\psibar_{x+\mu}\psi_{x}
-2\psibar_{x}\psi_{x}\right)\,.\nonumber
\end{eqnarray}
The right-handed fermions do not transform under the gauge group,
and for $r=0$ the fermion sector of the theory has a $G_L\times G_R$
symmetry, where only $G_L$ is local.\footnote{
  Obviously, for $r=0$ doublers are present; as in the standard QCD case
  \cite{ks}, they are absent for any fixed $r>0$.
  Similar statements apply to the chiral domain-wall fermion action
  of ref.~\cite{bd}.
}
For $r>0$, the Wilson term breaks this
symmetry to the diagonal subgroup, and thus breaks gauge, or (e)BRST,
invariance explicitly.  The theory is also invariant under a shift
symmetry on the right-handed spectator,
\begin{equation}
\label{shift}
\psi_R\to\psi_R+\e_R\,,
\end{equation}
which protects the theory
against an induced fermion mass term and any other relevant or marginal
operator involving the right-handed fermion \cite{gp}.  Note that the
spectators become free fields in the classical continuum limit.

The chiral Wilson action, Eq.~(\ref{lfermion}), is not the only acceptable one.
In fact, within the present framework, the familiar concept
of universality applies. This means that one can use any fermion
action with the correct classical continuum limit.
For a chiral fermion action based on domain-wall fermions
(which does not work without gauge fixing \cite{dwf}),
see ref.~\cite{bd}.

For $U_{x,\mu}=1$, obviously the fermions are not doubled, due
to the presence of the Wilson term.  However, one may worry that
doublers are generated dynamically, for instance,
if degrees of freedom exist
in the interacting theory which can form bound states with the
right-handed spectator fermions.\footnote{Or alternatively the
left-handed fermions may form screened bound states, decoupling from
the gauge fields \cite{gps}.}
Indeed, without gauge fixing, the gauge degrees of freedom
are not controlled by any small parameter, and all wavelengths
of these modes are equally important.  In an exactly gauge invariant
lattice theory (such as QCD in the commonly used lattice formulations)
this does not matter, because the gauge degrees of freedom decouple
from the physical degrees of freedom.  In chiral gauge theories,
however, each fermion field needs to contribute its ``share" of
the anomaly (even if the whole collection of fermion fields is
anomaly free) \cite{ks,bgspt}, and the regulated theory therefore
tends to break the gauge invariance by irrelevant terms.
This is certainly the case for the chiral Wilson action used here,
as well as for the chiral domain-wall fermion action of ref.~\cite{bd}.
It is therefore necessary to control the gauge degrees of freedom
in such a way as to decouple them in the continuum limit.

In our proposal, the lattice action contains kinetic terms for all four
polarizations of the gauge-field vector bosons.
Thus, all polarizations of the gauge field are controlled by the
gauge coupling $g$, including the longitudinal modes, which are
controlled by the gauge fixing. Moreover, as we will describe
below, the gauge
fixing on the lattice is done in such a way as to ensure
the existence of a unique classical vacuum. As a consequence, lattice
perturbation theory can be systematically developed.
Even though our theory is not
gauge invariant (hence not unitary) on the lattice, it is renormalizable.
As in ordinary lattice QCD with Wilson fermions, the Wilson term
remains a relevant operator for large momenta (of order $\pi/a$),
and doublers do not re-appear dynamically.  This was investigated
in more detail in perturbation theory in refs.~\cite{bgspt,bd}.
The interactions between the fermions and the longitudinal gauge fields
were also investigated numerically in the abelian case, finding again that
no doublers appear, and in fact, that there is good quantitative
agreement between these non-perturbative investigations and
perturbation theory \cite{bgsprl,bgs97,bd,bgsarenshoop}.\footnote{
The numerical investigation was for $G=U(1)$.  However, there is
little doubt that similar results would be obtained for $G=SU(N)$.}
Gauge fixing is essential for this conclusion, and it has indeed been
shown that without gauge fixing fermion doublers are
generated dynamically \cite{reviews,mg2000,fddyn}.  The formulation
of non-abelian chiral gauge theories developed here makes it
possible to extend these numerical tests to the non-abelian case
as well.

A finite number of counter terms is added in order
to adjust the theory such that (at least to any order in perturbation theory)
we recover the target continuum theory,
in which gauge degrees of freedom as well as spectator fermions
decouple, and unitarity is restored.
This is done by requiring the BRST identities
of the target theory to be satisfied in the continuum limit,
as already observed in ref.~\cite{rome}.
Standard power counting is used in order to organize the counter terms.
In particular, there is a finite number of them,
and only three have mass dimension less than four.

In a gauge-fixed theory with exact BRST invariance, the Gribov
problem makes it non-trivial to define the gauge-fixed theory
non-perturbatively, as we have seen in the preceding sections. However,
in an exactly gauge- (or BRST-) invariant lattice theory, it does not matter
around which of the copies of the classical vacuum
one develops perturbation theory.
The same is not necessarily true when the regularized
theory is not exactly gauge invariant.  In that case, different
Gribov copies may lead to different perturbative expansions around
them; the counter terms needed to regain gauge invariance in
perturbation theory around one copy may not be appropriate for some other copy,
and summing or averaging over all copies may thus not yield the
desired continuum limit.\footnote{For a detailed investigation
of this issue, see ref.~\cite{bglspd}.}  This problem is particularly
acute because of the fact that, for a simple lattice transcription
of the continuum Lorenz gauge, most Gribov copies
will correspond to ``rough" configurations, \ie\ most
of them are lattice artifacts.

We deal with this problem by choosing a discretization of the
gauge-fixing action such that the classical vacuum configuration $U_{x,\mu}=1$
is the unique minimum of the action,
which implies that lattice perturbation theory around
the classical vacuum is valid by construction \cite{sprd,gsplb}.
Thus, we will choose our gauge-fixing
lagrangian {\it not} just equal to the sum of $\cl^L_{gf}$ in
Eq.~(\ref{lgflattice}), and some simple lattice discretization
of $\sum_i(\partial_\mu A^i_\mu)^2$.  In fact,
our gauge-fixing action on the lattice
will break eBRST symmetry explicitly.  However, this is not
a ``new price" to pay, since the fermion sector of the theory already
breaks this symmetry explicitly, and counter terms are needed
anyway.

Before we review the construction of the gauge-fixing action on the
lattice, let us address a possibly confusing point.
We do choose to use a lattice gauge-fixing action
with a unique absolute minimum.
This, however, does not mean that ``continuum" Gribov
copies are removed from the theory.
In particular, for the classical vacuum $U_{x,\mu}=1$,
the lattice action of a continuum Gribov copy will be $O(a^2p^2)$,
with $p$ some physical scale.
In the continuum limit, such copies do contribute to the functional
integral.  To the extent that this class of copies plays a role
in the physics of non-abelian theories, they will do so in the
continuum limit of our lattice regularization.

We start from the equivariantly
gauge-fixed lattice Yang--Mills theory constructed in Sect~3.
In particular, we will choose the equivariant gauge-fixing
action to be $\cl_{G/H}$, defined in Eq.~(\ref{lnonab}).
Our $H$ gauge-fixing term will
be constructed as follows.  Defining
\begin{equation}
\label{latticehf}
\ca_{x,\mu}={1\over i}\sum_i T^i
\ \tr\left(T^i(U_{x,\mu}-U^\dagger_{x,\mu})\right)\,,
\end{equation}
our discretization of the continuum Lorenz gauge
$\tr(\partial_\mu A_\mu)^2/(\alpha g^2)$ will be
\begin{equation}
\label{latticeab}
\cl_H=\sum_i{1\over \alpha g^2}\
\tr\left(\partial^-_\mu \ca_{x,\mu}\right)^2\,,
\end{equation}
where $\partial^-_\mu$ is the backward difference operator,
\begin{equation}
\label{back}
\partial^-_\mu\phi_x=\phi_x-\phi_{x-\mu}\,.
\end{equation}
Note that we do not introduce ghosts for $H$.
As a consequence, on the lattice there is no BRST symmetry corresponding
to this part of the gauge fixing.  This is because, contrary to the
continuum, on the lattice the Faddeev--Popov operator corresponding
to $\cl_H$ is not a free lattice laplacian, but instead depends on
$U_{x,\mu}$ through irrelevant terms.  Therefore the identities
(\ref{ebrstidagain}, \ref{ononinv}) do not hold on the lattice
even in the lattice theory without fermions, and certainly not in the
theory with fermions (see also ref.~\cite{morebrst}).

Next, we wish to
add an irrelevant term $\cl_{irr}$ such that the total gauge-fixing
action $\sum_x(\cl_H+\cl_{G/H}+\cl_{irr})$ has an absolute minimum
at the perturbative vacuum configuration $U_{x,\mu}=1$.  Lattice
perturbation theory will then correspond to a systematic saddle-point
approximation around the perturbative vacuum.

Following ref.~\cite{gsplb}, first define
\begin{eqnarray}
\label{defs}
\cv_{x,\mu}&=&{1\over 2i}\left(U_{x,\mu}-U^\dagger_{x,\mu}\right)\,,\\
\cc_x&=&\sum_\mu\left(2-U_{x,\mu}-U^\dagger_{x-\mu,\mu}\right)\,,\nonumber\\
\cb_x&=&{1\over 4}\sum_\mu\left(\cv_{x,\mu}+\cv_{x-\mu,\mu}\right)^2\,.
\nonumber
\end{eqnarray}
Then $\cl_{irr}$ is defined as
\begin{equation}
\label{lirr}
\cl_{irr}={\tilde{r}\over g^2}\
\tr\left({1\over 2}(\cc^\dagger_x+\cc_x)+\cb_x\right)
\left({1\over 2}(\cc^\dagger_x+\cc_x)-\cb_x\right)\,,
\end{equation}
where $\tilde{r}>0$ is a new parameter.  Like the Wilson parameter $r$,
its precise value is not important, and we will take it to be of order one.
It was proved in ref.~\cite{gsplb} that $\cl_{irr}$ is non-negative
for all $U_{x,\mu}$ and that it vanishes only for $U_{x,\mu}=1$.\footnote{
  We proved it for $G=SU(N)$ or $SO(N)$, but expect it to
  be true for any simple $G$.
}
Since both $\cl_H$ and $\cl_{G/H}$ are manifestly non-negative, and
also vanish for $U_{x,\mu}=1$, our claim follows.
Putting everything together, on the lattice we take
\begin{equation}
\label{latticegf}
\cl_{gf,\,lattice}=\cl_H+\cl_{G/H}+\cl_{irr}+\cl_{ghost}
\end{equation}
as our gauge-fixing lagrangian.  Here $\cl_{ghost}$ refers to
the ghost part of $\cl^L_{gf}$, defined in Eq.~(\ref{lghost}),
and contains only the $\cg/\ch$ coset ghosts.

To gain insight into the role of $\cl_{irr}$, it is instructive
to consider the classical potential $V_{class}$ for constant,
commuting gauge fields, following refs.~\cite{sprd,gsplb}.
Because of the lack of gauge invariance, mass counter terms will be
needed for all gauge bosons, as we will discuss below in more detail.
Including mass (counter) terms, one has
\begin{eqnarray}
\label{classpot}
V_{class}={\tilde{r}\over 2 g^2}\ \left[\tr\left(\sum_\mu V_\mu^2\right)
\left(\sum_\nu V_\nu^4\right)+\dots\right]+\kappa\ \left[\tr\left(
V_\mu^2\right)+\dots\right]\,,
\end{eqnarray}
where at the classical level, we may choose
$\kappa_W=\kappa_A=\kappa$ (\cfdot\ Eq.~(\ref{mass})).
This potential exhibits a continuous phase transition between two phases.
For $\kappa>0$ we have a ferro-magnetic (FM) phase with $\svev{V_\mu}=0$.
For (small) $\kappa<0$ we have a directional ferro-magnetic phase (FMD)
in which the gauge field picks up an expectation value
\begin{eqnarray}
\label{vev}
\svev{V_\mu}=\pm\left({|\kappa|g^2\over 6 \tilde{r}}\right)^{1/4}\,,
\ \ \ \ \ {\rm all}\ \mu\,.
\end{eqnarray}
The continuum limit will be defined by approaching the critical line
from the FM ($\svev{V_\mu}=0$) side. Note that the lattice expectation
value (\ref{vev}) only breaks a discrete symmetry (hyper-cubic
rotations), and thus no Goldstone bosons occur inside the FMD phase.
This phase transition defines a novel universality class, with
a greatly enlarged symmetry at the critical point, at which both
gauge invariance and rotational invariance are recovered.\footnote{
  The names ``FM" and ``FMD"
  reflect the phase diagram obtained in the so-called
``reduced model limit,'' which constrains
the gauge field to pure-gauge configurations
$U_{x,\mu}=\phi_x \phi^\dagger_{x+\mu}$.
The FM phase is analytically connected to the Higgs phase in an
abelian theory, or to the Higgs-confinement phase in a non-abelian theory.
The FM-FMD phase transition is very different from the usual Higgs
transition, and separated from it by a multi-critical point
\cite{bgsphased,bglspd,bds}.
}
  Because of the existence of a unique classical
vacuum and the fact that we have a
consistent power counting, lattice perturbation
theory applies near this critical point.\footnote{
  On the FMD side, the classical vacua exhibit the (discrete) degeneracy
  dictated by the spontaneous breaking of hyper-cubic rotations.
  As usual, perturbation theory around one of these vacua is valid
  in the infinite-volume limit.
}
The existence of this novel critical point is therefore the key ingredient
making it possible to formulate a lattice gauge theory with
undoubled chiral fermions. The presence of the gauge-fixing sector
introduces a new direction to the phase diagram of the theory
(through the coupling $\tilde{r}/g^2$), thus giving access to this
critical point.  (Previous attempts with Wilson fermions without gauge fixing
failed because of the impossibility of reaching this novel
universality class.)

The irrelevant term, $\cl_{irr}$, represented by the $V^6$ term in
Eq.~(\ref{classpot}), stabilizes the classical potential.
The critical line separating the two phases with $\svev{V_\mu}=0$
and with $\svev{V_\mu}\ne 0$ corresponds to a vanishing curvature at
the origin, \ie\ to a vanishing gauge-field mass-squared.\footnote{
For more details on the nature of this critical point, see ref.~\cite{gsplb}.
}
Classically, the transition is at $\kappa=\kappa_c=0$.
In the full quantum theory,
the transition is expected to be continuous up to, possibly,
effects which are non-perturbatively small in the renormalized coupling
constant.  Without $\cl_{irr}$ the transition would be strongly first order
(\ie\ with a discontinuity in $\svev{V_\mu}$ of order $1/a$).
$U_\mu=1$ would not be the unique classical vacuum, and
perturbation theory around this vacuum would most likely
not correspond to a systematic expansion of the lattice theory.\footnote{
For some non-perturbative investigations of this issue in
an abelian theory, see ref.~\cite{bglspd}.
}

Once the existence of the critical point is secured by the irrelevant term,
$\cl_{irr}$ is indeed ``irrelevant" in that it does not affect
the long-distance physics near this critical point.\footnote{
  The Wilson term is irrelevant in precisely the same sense:
  {\it once} the doublers have been removed, it does not affect the
  long-distance physics of the one remaining relativistic fermion.
}
In summary, the fact that stability of the potential
at the critical point is obtained through the help of an
irrelevant operator implies
that that irrelevant operator does not occur in the renormalized
(continuum) lagrangian that governs the critical point,
which is the gauge-fixed
Yang--Mills theory described in sections 2 and 5, coupled
to left-handed chiral fermions.

We now come to the final step of our construction, the addition
of counter terms.  As mentioned already,
since (e)BRST is explicitly broken on the lattice,
counter terms will need to be added in order to be able to recover
the target chiral gauge theory in the continuum limit.  Since
our theory admits an expansion in the lattice coupling constant,
and satisfies the usual power counting rules, only counter terms
of engineering dimension less than or equal to four need to be
included.  While earlier work on counter terms in this context
was carried out \cite{rome,bglspd}, we will need to redo the job,
because of the different gauge condition, and the lack of shift
symmetry on the anti-ghost field.  The symmetries restricting
the number of counter terms are hyper-cubic symmetry,
$CP$ invariance, global $H$
invariance, the discrete subgroup $\St_N$ of $SU(N)$ introduced
in Sect.~2, flip symmetry, ghost-$SU(2)$ (which includes ghost-number)
symmetry, and the
shift symmetry (\ref{shift}) on the right-handed spectator fermions.

We begin with counter terms of dimension two.  There are three of
these: mass terms for the $A$ and $W$ fields as well as the ghost
field.  We can take
\begin{equation}
\label{mass}
\cl_{ct,\,d=2}=\kappa_A\ \tr(\ca_{x,\mu})^2-2\kappa_W\sum_i\
\tr\left(T^iU_{x,\mu}T^iU^\dagger_{x,\mu}\right)
+\kappa_C\ \tr(\Cbar_xC_x)\,.
\end{equation}
There are no dimension-three counter terms, since they are all
forbidden by the lattice symmetries.
In particular, a fermion mass is excluded by shift symmetry, Eq.~(\ref{shift}).
(When using the chiral domain-wall fermion action \cite{bd}
a similar conclusion applies in the limit of an infinite
fifth dimension where (global) chiral symmetry is recovered.)
In contrast, there are many marginal (dimension-four) counter terms.
Below, we will use continuum notation for all dimension-four counter terms;
they can easily be transcribed to the lattice, for instance by using
Eqs.~(\ref{latticehf}, \ref{cw}).  (It does not matter how exactly
the latticization is carried out, and more economical lattice versions
may exist.)  In the fermion sector, there are only two:
\begin{equation}
\label{fermion}
\cl_{ct,\,fermion}=\dg_A\psibar_L\gamma_\mu(iA_\mu)\psi_L
+\dg_W\psibar_L\gamma_\mu(iW_\mu)\psi_L\,.
\end{equation}
Note that our fermionic counter terms only involve the
left-handed fermion, since the right-handed spectator is again
protected by shift symmetry. As was shown in ref.~\cite{gp},
the right-handed spectators decouple automatically
if the theory has a continuum limit.  Because there is no global $SU(N)$
symmetry, the bare couplings $g_A$ and $g_W$ will have to be adjusted
separately in order to maintain universality of the renormalized
gauge coupling.

There is a large number of dimension-four counter terms involving
only the gauge fields or ghost fields, totaling 75,
including wave-function renormalizations for
$A$, $W$ and the ghosts.  For $N=2$ these are not all
independent, but for generic $N$ they are.  We list them in
Appendix D.

It would be prohibitively expensive to simultaneously tune
all these counter terms numerically in order to satisfy the desired
Slavnov--Taylor identities of the target theory in the continuum limit.
However, for small enough gauge coupling, this is
not needed.  The three relevant, mass counter terms will have to be tuned
non-perturbatively in order to move the lattice theory to the
critical point described by the target continuum theory.  In other
words, the three mass terms of Eq.~(\ref{mass}) need to be adjusted
non-perturbatively toward infinite correlation lengths for the
degrees of freedom associated with the fields $A$, $W$ and the ghosts.
While the ghost fields are not physical, they will have to remain
(perturbatively) massless if the continuum theory is to be
unitary.

In contrast, all dimension-four counter terms can be estimated with the
help of lattice perturbation theory, because our lattice theory has
been constructed such that perturbation theory is a systematic
approximation.  If tree-level precision is sufficient, this implies
that they can all be omitted.  If not, a one-loop calculation
of these counter terms should improve the situation.  While this
is not a simple calculation, it can be done analytically, and
parametrically as a function of the free parameters $g$, $\xi$
and $\alpha$.
Alternatively, one may imagine a numerical determination, where
first $\k_{A,W,C}$ are determined without any other counter terms
present.  After that, one may determine each of the dimension-four
counter terms by considering the appropriate correlations functions,
\ie\ the ones to which they would contribute to lowest non-trivial
order.  Obviously, this still relies on the validity of perturbation
theory, but may help in side-stepping questions as to which
value of the coupling constant should be used in the one-loop
expressions for these counter terms in a particular numerical computation.

\vspace{5ex}
\noindent {\large\bf 7.~Conclusion}
\secteq{7}
\vspace{3ex}

We believe that this work represents major progress in the
non-perturbative construction of non-abelian chiral gauge
theories.  While we already described in some detail what
our construction does and does not accomplish in the
Introduction, let us summarize what has been gained.

The key ingredients of the lattice chiral gauge theories constructed
in this paper are the use of a renormalizable gauge, as well as
the existence of a unique classical vacuum.
This leads to the existence of a novel type of critical point,
with a systematic weak-coupling expansion, and
the target continuum chiral gauge theory is recovered to all orders
in this expansion at this critical point.

It is instructive to compare our lattice construction with the standard
perturbative treatment of chiral gauge theories in the continuum,
for instance through the use of dimensional regularization.
The continuum-regularized chiral gauge theory shares with our lattice theory
the following crucial features:
1) Both regularized theories do not preserve the chiral gauge invariance.
2) As a result, counter terms have to
be added to recover gauge invariance of the renormalized theory.
3) The regularized theory is in fact renormalizable {\it without}
relying on gauge invariance, thanks to the existence of kinetic terms
for all four polarization, where the longitudinal kinetic term
is provided by the gauge-fixing action.

The main difference is that,
in the continuum, it is simply {\it assumed} that some
non-perturbative theory exists for which the perturbative
expansion is valid.  The question as to whether a non-perturbative
formulation actually exists with a critical point which is indeed described by
renormalized perturbation theory is not even asked.  On the lattice,
a theory with the appropriate critical point will have to be constructed
explicitly, and this is what we did in the present paper.
In the lattice construction, the use of a renormalizable
gauge and the uniqueness of the classical vacuum are
equally important, and independent, ingredients.

The existence of a valid perturbative expansion, or,
in other words, the fact that the universality class is known,
is a feature common to standard Lattice QCD and to our construction.
The difference is that standard Lattice QCD is exactly gauge invariant,
and gauge fixing is an extraneous device needed only to set up
weak-coupling perturbation theory;
in contrast, in our construction the regularized theory is not
gauge invariant, and renormalizability is maintained because
the gauge-fixed
lattice action explicitly contains longitudinal kinetic terms
for all gauge bosons.
{\it The specific gauge fixing we adopted is thus part of the very
definition of the theory.}

Of course, this does not mean that we now know that unitary,
Lorentz-invariant chiral gauge theories with gauge group
$G=SU(N)$ exist for anomaly-free fermion representations.
For example, the $SU(2)$ theory with one fundamental-representation
Weyl fermion does not exist because of a non-perturbative
obstruction \cite{witten,bc}.  However, whether any other
non-perturbative obstruction exists for a certain gauge
group and fermion content is a dynamical question
in the context of our construction, and can in principle
be studied by non-perturbative analytical or numerical techniques.

As we explained in more detail in Sect.~6 (see in particular the
discussion around Eq.~(\ref{classpot})),
the single most important dynamical feature of the critical point is
that the chiral nature of the fermion spectrum is maintained
non-perturbatively \cite{bgsprl}.\footnote{
  For an explanation on how the construction by-passes the Nielsen-Ninomiya
  theorem \cite{nn} see ref.~\cite{bgs97}.
}
Since the familiar notion of universality applies,
this should be true not only for the chiral Wilson action used by us,
but in fact for any lattice fermion action
with the correct classical continuum limit.
For the chiral domain-wall fermion action,
this was demonstrated in ref.~\cite{bd}.
Thus, fermion doublers are {\it not} generated dynamically,
and no fermion masses can
occur without a dynamical breaking of the gauge group.
Scenarios for the dynamical symmetry breaking of a chiral gauge theory,
or for light composite fermions \cite{tumbl},
can in principle be studied within our construction.

It is interesting to contrast this state of affairs with the
approach of ref.~\cite{mlnonab}.  In that approach,
the goal is to construct non-abelian lattice chiral gauge theories with
exact gauge invariance. Like in Lattice QCD with exact
gauge invariance, if this goal would be reached, there would be
no need to worry about the back reaction of the gauge degrees
of freedom on the fermion spectrum (or on any other physical
degree of freedom); the unphysical degrees of freedom
would decouple in the regulated theory,
due to the exact gauge invariance.  Obviously,
this approach necessitates a complete classification of all possible
non-perturbative obstructions. This problem was
formulated in terms of suitable integrability conditions in ref.~\cite{mlnonab},
and was solved perturbatively in ref.~\cite{mlpert}.
A non-perturbative solution of the integrability
conditions constitutes a much harder problem,
and indeed, to date no solution is known.\footnote{
  In the abelian case a complete classification exists,
  and exact gauge invariance can be established provided
  the fermion spectrum satisfies one new condition
  apart from the usual anomaly-cancellation condition \cite{mlabelian}.
  See also ref.~\cite{mg2000}.
}
In other words, it has not been established
that the necessary critical point exists within this approach
for the non-abelian case.

In our construction, we do not insist on exact gauge invariance
on the lattice.  Instead, because of the gauge fixing,
{\it all} degrees of freedom are under dynamical control.
The continuum limit of an asymptotically-free theory
corresponds to a vanishing bare coupling constant,
and the validity of the weak-coupling expansion means that
the elementary degrees of freedom are those, and only those, that occur
at tree level in perturbation theory.
In other words,  the target (chiral) gauge theory, whose
particle content and interactions may be read off from
the {\it classical} continuum limit of the lattice action,
is indeed realized in the {\it quantum} continuum limit of the lattice theory.
The lattice dynamics does not generate any
new light degrees of freedom not already contained in
this target continuum theory \cite{bgsprl,bgs97,bglspd}.
Also, the unphysical degrees of freedom
of the target gauge-fixed theory decouple in the continuum limit
after the adjustment of a finite number of counter terms
(at least) to all orders in perturbation theory.
Because perturbation theory is reliable at the lattice scale,
any non-perturbative obstruction to this
conclusion can only originate in some infra-red mechanism
contained in the theory.  This is precisely what makes
it interesting to apply the construction proposed in this
paper to the study of non-abelian chiral gauge theories.

The approach of ref.~\cite{mlnonab} has
yielded a ``by-product" which is a
gauge-invariant lattice weak-coupling expansion
for anomaly-free chiral gauge theories \cite{mlpert} (see also
refs.~\cite{suzuki,banetal}).
It is, as already mentioned above, an open question whether a critical
point exists which is controlled by this expansion,
simply because the underlying non-perturbative lattice theory
is not (yet) known.  The perturbative solution of ref.~\cite{mlpert}
involves the adjustment of an infinite number of irrelevant operators,
in order to enforce exact gauge invariance on the lattice to any
given order. It is unlikely that universality holds in that case.
In the gauge-fixing approach, universality applies, and
the correct continuum limit is obtained to all orders
after the adjustment of a finite number of (relevant and marginal)
counter terms.
The gauge-fixing approach also goes beyond this in providing
a fully non-perturbative lattice theory.
The actual number of counter terms is undeniably large.
However, in Sect.~6, we have
argued that this is unlikely to be a very severe obstacle,
because most can be reliably calculated in low-order
perturbation theory.

We believe that our results are interesting for the case
of pure Yang--Mills (or vector-like) theories as well.
We demonstrated in Sect.~4
that the equivariantly gauge-fixed lattice Yang--Mills theory
is rigorously equivalent to the non-gauge-fixed and thus
gauge-invariant theory.  Among other things, it follows
that the equivariantly gauge-fixed theory is unitary, because
the gauge-invariant theory is.  It would be interesting
to see whether this conclusion can be extended non-perturbatively
to the remaining maximal abelian group $H$, as we did
to all orders in perturbation theory in Sect.~5.
It is possible that the idea proposed in ref.~\cite{testa}
can be extended to our case.  The reason that this is
not trivial, however, is the ``entanglement" of the
abelian and non-abelian degrees of freedom.

The non-perturbative study of chiral gauge theories following
the approach outlined in this paper is not an easy task.
With an assortment of fermion and ghost determinants, numerical
investigations will certainly be very demanding.
The theory contains, apart from a complex fermion determinant,
also a ghost determinant (the four-ghost interaction terms
can be transformed into bilinear terms with the help of
bosonic auxiliary fields).  We envisage to begin with a study
of the non-perturbatively gauge-fixed Yang--Mills theory, with
no fermions.  This should be interesting by itself.  Moreover,
it is likely that much can
be learned about the phase diagram of lattice theories
as constructed here by a combination of weak- and strong-coupling
analytic methods;  work in this direction is in progress.
Concerning the fermions, it should prove
useful to begin with a study of the phase of the fermion
determinant on an ensemble of quenched configurations.
To the extent that chiral gauge theories are qualitatively
different from vector-like gauge theories, this difference
should originate in the phase of the determinant.  As argued in
ref.~\cite{bgsprl}, it is also sensible to check the absence of
fermion doublers for non-abelian gauge groups numerically
in the quenched theory, as was done there for $G=U(1)$.

We should emphasize however, that the construction outlined
here now makes such investigations at least in principle possible.
We have developed a first complete non-perturbative formulation
of non-abelian chiral gauge theories, and we have provided
what we believe to be compelling evidence that,
if a certain asymptotically-free chiral gauge theory exists,
it can be studied non-perturbatively using this formulation.

\vspace{5ex}
\noindent {\large\bf Acknowledgements}
\vspace{3ex}

We thank Aharon Casher, Jeff Greensite and Pierre van Baal
for useful discussions.
We would also like to thank the Institute for Nuclear Theory at the
University of Washington, Seattle, for hospitality.
YS thanks the Physics Department at San Francisco State University
for hospitality.
YS is supported by the Israel Science Foundation under grant
222/02-1. MG is supported in part by the US Department of Energy.

\vspace{5ex}
\noindent {\large\bf Appendix A}
\secteq{A}
\vspace{3ex}

Here we show that the single-site partition function $Z_{ghost}(1,0)$
defined by the last row of Eq.~(\ref{finalc}) is non-zero.
Introducing auxiliary fields $\rho_i$ we write
\begin{eqnarray}
  \int dC\, d\Cbar\ \exp\left[\tr(\Xt^2)\right]
  &=&
  \int dC\, d\Cbar\, d\rho\ \exp\left[
  -\sum_i
  \left( {1\over 2}\, \r_i^2 + \r_i f_{i\a\b}\Cbar_\a C_\b \right)
  \right]\,.
\NON
  &=& \int d\rho\ \exp\left(-{1\over 2}\sum_i \r_i^2 \right)
  \det(\rho_i f_{i\alpha\beta}) \,.
\label{rhoagain}
\end{eqnarray}
The matrix $\r_{\a\b} \equiv \r_i f_{i\a\b}$
is real, anti-symmetric, and even dimensional.
Hence its determinant never changes sign. This implies that the
single-site ghost determinant is non-negative.\footnote{
  We assume a suitable sign convention for
the Grassmann integration measure.
}

It remains to prove that the ghost determinant is non-zero
for {\it some} choice of $\rho_i$. This is where we will use
that $H$ is the maximal abelian subgroup of $G=SU(N)$.\footnote{
  The argument generalizes
  trivially to larger subgroups $H'$ such that $H \subset H'$.
  More generally, it was suggested that $Z_{ghost}$ will be non-zero provided
  that the Euler characteristic of the coset manifold $\cg/\ch$
  is non-zero \cite{schaden}.
}
For definiteness, we choose $\rho_i \ne 0$ to be
proportional to that linear combination which couples
to the following linear combination of diagonal generators
(in arbitrary normalization)
\begin{equation}
  T' = diag(1,2,\ldots,N-1,-N(N-1)/2) \,.
\end{equation}
It is easy to see that none of the off-diagonal $SU(N)$
generators commute with $T'$. Moreover the resulting matrix $\r_{\a\b}$
is skew-diagonal in the basis of off-diagonal generators introduced
in Eq.~(\ref{offdg}). For each pair $T^k_{AB}$, $k=1,2$, we obtain
a non-zero anti-symmetric two-by-two block, which implies that
$\det(\r_{\a\b})$ is non-zero in this case.

\vspace{5ex}
\noindent {\large\bf Appendix B}
\secteq{B}
\vspace{3ex}

In this Appendix, we comment on the Feynman rules
in the formalism in which the auxiliary fields $b$ and $\beta$
are kept.  The only part which is not completely straightforward
is that of the tree-level two-point functions for $W_\mu$ and
$A_\mu$.  The quadratic part of the action for $W_\mu$ and $b$
(the argument for $A_\mu$ and $\beta$ is similar) is,
after rescaling $W_\mu \to g W_\mu$, $b \to b/g$,
\begin{equation}
\label{lquad}
\cs_{quad}=
{1\over 2} \int d^4x\ \left(
(\partial_\mu W_\nu)^2-(\partial_\mu W_\mu)^2
-2ib\partial_\mu W_\mu+\xi b^2\right)\,,
\end{equation}
from Eqs.~(\ref{YM}) and~(\ref{lgfdetail}).  In momentum space,
this can be written as
\begin{equation}
\label{lquadmom}
\cs_{quad}=  {1\over 2} \int {d^4p\over (2\p)^4}\;
\pmatrix{W_\mu(-p) & b(-p)}
\pmatrix{p^2\delta_{\mu\nu}-p_\mu p_\nu & -p_\mu \cr
p_\nu & \xi}\pmatrix{W_\nu(p) \cr b(p)}\,,
\end{equation}
from which it follows that
\begin{eqnarray}
\label{props}
\svev{ W_\mu(p)W_\nu(q)}&=&{1\over p^2}
\left(\delta_{\mu\nu}+(\xi-1){p_\mu p_\nu\over p^2}\right)
\delta(p+q)\,,\\
\svev{ W_\mu(p)b(q)}&=&
-\svev{ b(p)W_\mu(q)}\ =\ {p_\mu \over p^2}\delta(p+q)\,,
\nonumber \\
\svev{ b(p)b(q)}&=&0\,. \nonumber
\end{eqnarray}
As expected, the $\svev{ W_\mu W_\nu}$ propagator is the same as in the
more familiar formulation where the auxiliary field is integrated out.

\vspace{5ex}
\noindent {\large\bf Appendix C}
\secteq{C}
\vspace{3ex}

Here we generalize the proof that the scattering matrix is independent
of the gauge parameter(s) to include gauge bosons
on the external legs. This requires an additional argument since
the (e)BRST transformation of a gauge boson contains a term
linear in a ghost field. Explicitly,
$s (g W_\mu^\alpha) = \partial_\m C^\alpha + O(g)$,
$s_H (g A_\mu^i) = \partial_\mu \chi^i$. For definiteness, consider
$\co = W_\nu^\alpha(k) \co_1$ where $k$ is the $W$'s momentum
and $\co_1$ is a product of fermion fields (the argument generalizes
trivially to scattering amplitudes involving several gauge bosons).
The right-hand side of Eq.~(\ref{ddxi}) then contains a term linear in
the $C$ ghost:
\begin{equation}
  k_\nu \vev{C^\alpha(k) \co_1 \int d^4x\, s\sbar(\Cbar C)} \,.
\label{Cterm}
\end{equation}
In order to go on-shell we first multiply both sides of Eq.~(\ref{ddxi})
by the $W$'s equation-of-motion operator,
contracted with a (normalized) transverse polarization vector
\begin{equation}
  \epsilon^{(n)}_\mu ( k^2 g^{\mu\nu} + (1/\xi - 1) k^\mu k^\nu)
  = \epsilon^{(n)\nu} k^2 \,.
\label{eom}
\end{equation}
The transverse polarizations are defined by the conditions
$\epsilon^{(n)}_4=0$ and ${\vec\epsilon}\, {}^{(n)}\cdot\vec{k}=0$.
(The transverse polarizations are well defined
already before we take the momentum on-shell.
Also notice that the contraction in Eq.~(\ref{eom}) is independent of $\xi$,
and thus commutes with $d/d\xi$.)
Since $\epsilon^{(n)}_\mu k^\mu=0$, the product of expressions~(\ref{Cterm})
and~(\ref{eom}) vanishes (even before we take the $W$'s momentum on shell).
A similar reasoning applies to scattering
amplitudes involving an $H$-subgroup gauge boson $A_\mu^i$.
This completes the proof that all scattering amplitudes are independent
of the gauge parameters $\xi$ and $\alpha$.

\vspace{5ex}
\noindent {\large\bf Appendix D}
\secteq{D}
\vspace{3ex}

In this Appendix we list all four-dimensional counter terms
constructed from the fields $A$, $W$, $C$ and $\Cbar$,
which we will generically label as $\Phi_p$, with $p$ a label
running over $A$, $W$, $C$ and $\Cbar$ (see Eq.~(\ref{fermion}) for the
four-dimensional counter terms involving fermions). As mentioned in the
text, we may construct the counter terms in the continuum, and
then use a suitable discretization which does not violate any
of the symmetries present on the lattice.
These symmetries are global $H$ invariance,
the discrete group $\St_N\in SU(N)$, flip symmetry,
ghost-$SU(2)$ (which includes ghost number), hyper-cubic symmetry
and $CP$ invariance.
One generates all
possibilities by taking traces over products over $\Phi_p$,
with or without derivatives, up to dimension four.
Since all $\Phi_p$ are traceless, there are terms containing
one or two traces only.  One then imposes other symmetries.
For instance, for each $\Cbar$ there has to be a $C$, and
some possibilities have to be chosen in linear combinations
which are invariant under flip and/or ghost-$SU(2)$ symmetry.

In addition, since projecting on the subalgebra $\ch$ or the coset
$\cg/\ch$ is invariant under $\St_N$ and $H$, each trace over
a product of fields can generate new terms through
replacements like
\begin{equation}
\label{hproj}
\tr(\Phi_1\Phi_2\Phi_3\Phi_4)\to
\tr(\Phi_1\Phi_2\Phi_3\Phi_4)+
\tr\left(\left(\Phi_1\Phi_2\right)_\ch\left(\Phi_3\Phi_4\right)_\ch\right)
+\tr\left(\left(\Phi_2\Phi_3\right)_\ch\left(\Phi_4\Phi_1\right)_\ch\right)
\,,
\end{equation}
where each term of course comes with an arbitrary coefficient.
In general, the product of two fields can be written as
\begin{equation}
\label{prod}
\Phi_1\Phi_2=\left(\Phi_1\Phi_2\right)_1+\left(\Phi_1\Phi_2\right)_\ch
+\left(\Phi_1\Phi_2\right)_{\cg/\ch}\,,
\end{equation}
where with the subscript ``$1$" we indicate the part proportional
to the identity matrix.  We can thus pick any three of these
products as independent, and construct counter terms out of them.
Here we will choose $\Phi_1\Phi_2$, $\left(\Phi_1\Phi_2\right)_1$
and $\left(\Phi_1\Phi_2\right)_\ch$.
The term of the form $\left(\Phi_1\Phi_2\right)_1$ can be
written in terms of $\tr(\Phi_1\Phi_2)$, and thus leads to a
term with two traces.

The projection onto $\ch$ can be understood in a slightly different
way.  Consider the $H$-invariant (which is also invariant under
$\St_N$, as we showed in Sect.~2)
\begin{equation}
\label{hinv}
\sum_i\ \tr(T^iXT^iY)\,,
\end{equation}
for any $N\times N$ matrices $X$ and $Y$.  In the orthogonal basis
defined by $\tr(T^iT^j)=\half\delta_{ij}$, it is possible to prove that
\begin{equation}
\label{ticompl}
\sum_i (T^i)_{AB}(T^i)_{CD}={1\over 2}\delta_{AB}\delta_{AC}\delta_{AD}
-{1\over 2N}\delta_{AB}\delta_{CD}\,.
\end{equation}
Using this relation, one obtains
\begin{equation}
\label{eq}
\sum_i\ \tr(T^iXT^iY)={1\over 2}\sum_A X_{AA}Y_{AA}-{1\over 2N}
\ \tr(XY)\,.
\end{equation}
For the $\ch$-projected matrices $X_\ch=2T^i\ \tr(T^iX)$ and $Y_\ch$ one has
\begin{equation}
\label{hprojprod}
\tr(X_\ch Y_\ch)=\sum_A X_{AA}Y_{AA}-{1\over N}\ \tr(X)\ \tr(Y)\,,
\end{equation}
hence
\begin{equation}
\label{relation}
\tr(X_\ch Y_\ch)=2\sum_i\ \tr(T^iXT^iY)+{1\over N}\ \tr(XY)
-{1\over N}\ \tr(X)\ \tr(Y)\,.
\end{equation}
Applying this for example to the pre-potential in Eq.~(\ref{lgflattice})
gives
\begin{equation}
\label{ppdiff}
2\sum_i\ \tr\left(T^iU_{x,\mu}T^iU^\dagger_{x,\mu}\right)
=\tr\left((U_{x,\mu})_\ch(U^\dagger_{x,\mu})_\ch\right)
+{1\over N}\ \tr(U_{x,\mu})\ \tr(U^\dagger_{x,\mu})-1\,.
\end{equation}

We list all counter terms below with no explicit couplings,
but arbitrary real coupling constants are implied in front of
each operator.  If two operators are related by flip (or ghost-$SU(2)$)
symmetry, we will include them together between parentheses.

Counter terms with two fields (and thus two derivatives) are
\begin{eqnarray}
\label{cttwo}
\cl_{c.t.\,2}&=&\tr(\partial_\mu W_\mu\partial_\nu W_\nu)
+\tr(\partial_\mu W_\nu\partial_\mu W_\nu)
+\tr(\partial_\mu W_\mu\partial_\mu W_\mu)\\
&&+\tr(\partial_\mu A_\mu\partial_\nu A_\nu)
+\tr(\partial_\mu A_\nu\partial_\mu A_\nu)
+\tr(\partial_\mu A_\mu\partial_\mu A_\mu)\nonumber\\
&&+\tr(\partial_\mu\Cbar\partial_\mu C)\,.\nonumber
\end{eqnarray}
Summation over {\it all} repeated space-time indices is assumed,
so, \eg $\tr(\partial_\mu W_\mu\partial_\mu W_\mu)$
is a shorthand for $\sum_\mu\tr(\partial_\mu W_\mu\partial_\mu W_\mu)$.
Counter terms with three fields (and thus one derivative) are
\begin{eqnarray}
\label{ctthree}
\cl_{c.t.\,3}&=&
\tr(\partial_\mu A_\mu A_\nu A_\nu)+\tr(\partial_\mu A_\nu A_\mu A_\nu)
+\tr(\partial_\mu A_\mu A_\mu A_\mu)\\
&&+\tr(\partial_\mu W_\mu W_\nu W_\nu)+\tr(\partial_\mu W_\nu\{W_\mu,W_\nu\})
+i\ \tr(\partial_\mu W_\nu[W_\mu,W_\nu])\nonumber\\
&&+\tr(\partial_\mu W_\mu W_\mu W_\mu)
\nonumber\\
&&+\tr(A_\mu\{\partial_\mu W_\nu,W_\nu\})+
i\ \tr(A_\mu[\partial_\mu W_\nu,W_\nu])\nonumber\\
&&+\tr(A_\mu\{\partial_\nu W_\mu,W_\nu\})+
i\ \tr(A_\mu[\partial_\nu W_\mu,W_\nu])\nonumber\\
&&+\tr(A_\mu\{\partial_\nu W_\nu,W_\mu\})+
i\ \tr(A_\mu[\partial_\nu W_\nu,W_\mu])\nonumber\\
&&+\tr(A_\mu\{\partial_\mu W_\mu,W_\mu\})+
i\ \tr(A_\mu[\partial_\mu W_\mu,W_\mu])\nonumber\\
&&+i\ \tr\left(A_\mu(\{\partial_\mu\Cbar,C\}-\{\partial_\mu C,\Cbar\})\right)
+\tr\left(A_\mu([\partial_\mu\Cbar,C]-[\partial_\mu C,\Cbar])\right)
\nonumber\\
&&+i\ \tr\left(W_\mu(\{\partial_\mu\Cbar,C\}-\{\partial_\mu C,\Cbar\})\right)
+\tr\left(W_\mu([\partial_\mu\Cbar,C]-[\partial_\mu C,\Cbar])\right)
\,.\nonumber
\end{eqnarray}
The last four terms are examples of linear combinations invariant
under flip and ghost-$SU(2)$ symmetry.
For the terms with four fields, we get terms with two traces,
\begin{eqnarray}
\label{ctfourtt}
\cl_{c.t.\,22}&=&
\tr(A_\mu A_\mu)\ \tr(A_\nu A_\nu)+\tr(A_\mu A_\nu)\ \tr(A_\mu A_\nu)
+\tr(A_\mu A_\mu)\ \tr(A_\mu A_\mu)\\
&&+\tr(W_\mu W_\mu)\ \tr(W_\nu W_\nu)+\tr(W_\mu W_\nu)\ \tr(W_\mu W_\nu)
+\tr(W_\mu W_\mu)\ \tr(W_\mu W_\mu)\nonumber\\
&&+\tr(A_\mu A_\mu)\ \tr(W_\nu W_\nu)+\tr(A_\mu A_\nu)\ \tr(W_\mu W_\nu)
+\tr(A_\mu A_\mu)\ \tr(W_\mu W_\mu)\nonumber\\
&&+\tr(\Cbar C)\ \tr(A_\mu A_\mu)+\tr(\Cbar C)\ \tr(W_\mu W_\mu)
+\tr(\Cbar W_\mu)\ \tr(CW_\mu)\,,\nonumber
\end{eqnarray}
and, finally, terms with only one trace,
\begin{eqnarray}
\label{ctfourot}
\cl_{c.t.\,4}&=&
\tr(A_\mu A_\mu A_\nu A_\nu)+\tr(A_\mu A_\mu A_\mu A_\mu)\\
&&+\tr(\{A_\mu,W_\nu\}\{A_\mu,W_\nu\})+\tr([A_\mu,W_\nu][A_\mu,W_\nu])
\nonumber\\
&&+\tr(\{A_\mu,W_\mu\}\{A_\nu,W_\nu\})+\tr([A_\mu,W_\mu][A_\nu,W_\nu])
+i\ \tr([A_\mu,W_\mu]\{A_\nu,W_\nu\})\nonumber\\
&&+\tr(\{A_\mu,W_\mu\}\{A_\mu,W_\mu\})+\tr([A_\mu,W_\mu][A_\mu,W_\mu])
\nonumber\\
&&+\tr(\{A_\mu,W_\nu\}\{W_\mu,W_\nu\})+\tr([A_\mu,W_\nu][W_\mu,W_\nu])
\nonumber\\
&&+i\ \tr([A_\mu,W_\nu]\{W_\mu,W_\nu\})
+\tr(A_\mu W_\mu W_\mu W_\mu)
\nonumber\\
&&+\tr(W_\mu W_\mu W_\nu W_\nu)+\tr(W_\mu W_\nu W_\mu W_\nu)
+\tr(W_\mu W_\mu W_\mu W_\mu)\nonumber\\
&&+\tr((W_\mu W_\mu)_\ch (W_\nu W_\nu)_\ch)
+\tr((W_\mu W_\mu)_\ch (W_\mu W_\mu)_\ch)\nonumber\\
&&+\tr([W_\mu, W_\nu]_\ch [W_\mu, W_\nu]_\ch)
+\tr(\{W_\mu, W_\nu\}_\ch \{W_\mu, W_\nu\}_\ch)\nonumber\\
&&+\tr(\{\Cbar,W_\mu\}\{C,W_\mu\})+\tr([\Cbar,W_\mu][C,W_\mu])\nonumber\\
&&+\tr(\{\Cbar,A_\mu\}\{C,A_\mu\})+\tr([\Cbar,A_\mu][C,A_\mu])\nonumber\\
&&+\tr\left(\{\Cbar,W_\mu\}\{C,A_\mu\}+
\{\Cbar,A_\mu\}\{C,W_\mu\}\right)\nonumber\\
&&+i\ \tr\left(\{\Cbar,W_\mu\}[C,A_\mu]+
[\Cbar,A_\mu]\{C,W_\mu\}\right)\nonumber\\
&&+\tr([\Cbar,W_\mu][C,A_\mu]+[\Cbar,A_\mu][C,W_\mu])\nonumber\\
&&+\tr(\{W_\mu,\Cbar\}_\ch\{W_\mu,C\}_\ch)
+\tr([W_\mu,\Cbar]_\ch[W_\mu,C]_\ch)
\nonumber\\
&&+\tr\left(\{W_\mu,\Cbar\}_\ch[W_\mu,C]_\ch
+[W_\mu,\Cbar]_\ch\{W_\mu,C\}_\ch\right)\nonumber\\
&&+\tr(\Cbar^2 C^2)+\tr([\Cbar,C]_\ch[\Cbar,C]_\ch)\nonumber\\
&&+\tr\left((\Cbar^2)_\ch(C^2)_\ch
-(1/4)\{\Cbar,C\}_\ch\{\Cbar,C\}_\ch\right)\,.\nonumber
\end{eqnarray}
The operator $\tr(\Cbar C\Cbar C)$ is odd under flip symmetry, and
thus excluded.  Note that
\begin{equation}
\label{trace}
\tr([X,Z]\{Y,Z\})=-\ \tr(\{X,Z\}[Y,Z])=-\ \tr([X,Y]Z^2)\,,
\end{equation}
for any $X$, $Y$ and $Z$ (if both $X$ and $Y$ are anti-commuting,
$[X,Y]$ in the last expression should be replaced by $\{X,Y\}$).
In particular, if $X=Y$, this trace vanishes.
We used this relation to eliminate a number of terms in $\cl_{c.t.\,4}$.
The only new constraint imposed by extending ghost-number symmetry
to ghost-$SU(2)$ arises from the fact
that the latter mixes the two terms on the last row
of Eq.~(\ref{ctfourot}) (compare Eq.~(\ref{gh4b})).
Note that there are no counter terms involving
the Levi--Civita tensor $\e_{\mu\nu\rho\sigma}$ because of $CP$
invariance.

\vspace{5ex}

\end{document}